\documentclass[twocolumn,traditabstract]{aa}

\usepackage{natbib}
\usepackage{aas_macros}

\newcommand{\bibfont}{\tiny}
\usepackage{graphicx}
\usepackage[T1]{fontenc}
\usepackage{amsmath}
\usepackage{amssymb}
\usepackage{amsmath}
\usepackage{graphicx}
\usepackage{amsmath}
\usepackage{amssymb}
\usepackage{amsfonts}
\usepackage{graphicx,epsfig,graphics}
\usepackage{aas_macros}
\usepackage{graphics}
\usepackage{color}
\usepackage{epstopdf}

\def\be{\begin{eqnarray}}
\def\ee{\end{eqnarray}}
\graphicspath{{}}

\newcommand{\jbc}[1]{#1}

\titlerunning{Joint Planck and WMAP CMB Map Reconstruction}
\authorrunning{Bobin et al}

\title{Joint Planck and WMAP CMB Map Reconstruction}
\author{ \hspace{0.25in} J. Bobin\inst{1}, F. Sureau\inst{1}, J.-L. Starck\inst{1},  A. Rassat\inst{2}, P. Paykari\inst{1}}
\institute{$^1$ Laboratoire AIM, UMR CEA-CNRS-Paris 7, Irfu, SAp/SEDI, Service d'Astrophysique, CEA Saclay, F-91191 GIF-SUR-YVETTE CEDEX, France.\\
$^2$ Laboratoire d'astrophysique, Ecole Polytechnique F\'ed\'erale de Lausanne (EPFL), Observatoire de Sauverny, CH-1290, Versoix, Switzerland.}

\begin{document}
 
%%%%%%%%%%%%%%%%%%%%%%%%%%%%%%%%%%%%%%%%%%%%%%%%%%%%%%%%%%%%%%%%%%
%\abstract{We estimate a low-foreground map and power spectrum of Cosmic Microwave %background (CMB) from WMAP $7$-year data. 
%The method used is based on sparsity maximization algorithm named Local-Generalized Morphological Component Analysis (LGMCA), 
%implemented on a decomposition of the observations onto a wavelet domino. }

\abstract{We present a novel estimate of the cosmological microwave background (CMB) map by combining the two latest full-sky microwave surveys: WMAP nine-year and Planck PR1. The joint processing  
 benefits from a recently introduced component separation method coined { ``local}-generalized morphological component analysis'' (LGMCA) based on the sparse distribution of the foregrounds in the wavelet domain. \jbc{The proposed estimation procedure takes advantage of the IRIS $100 \mu m$ as an extra observation on the galactic center for enhanced dust removal.}
We show that this new CMB map presents several interesting aspects: i) it is a full sky map without using 
any inpainting or interpolating method, ii) foreground contamination is very low, iii) the Galactic center  is very clean, with especially low dust contamination as measured by the cross-correlation between the estimated CMB map and the IRIS $100 \mu m$ map, and iv) it is free of thermal SZ contamination.}
% The joint processing of the WMAP and Planck data allows for the estimation of the first true full-sky estimation of the CMB without any inpainted or interpolated pixels with low foreground contamination. It further leads to a very clean estimate of the Galactic center with especially low dust contamination as measured by the cross-correlation between the estimated CMB map and the IRIS $100 \mu m$ map. We finally show that the proposed CMB map is free of SZ contamination while being consistent with the Planck PR1 results otherwise.}

%%%%%%%%%%%%%%%%%%%%%%%%%%%%%%%%%%%%%%%%%%%%%%%%%%%%%%%%%%%%%%%%%%
\keywords{Cosmology : Cosmic Microwave Background, Methods : Data Analysis, Methods : Statistical}
\date{Received -; accepted -}
\maketitle

%%%%%%%%%%%%%%%%%%%%%%%%%%%%%%%%%%%%%%%%%%%%%%%%%%%%%%%%%%%%%%%%%%
\section{Introduction}
 
The Cosmic Microwave Background (CMB) is a snapshot of the state of the Universe at the time of recombination. It provides information about the primordial Universe and its evolution to the current state. Our current understanding of our Universe is heavily based on measurements of the CMB radiation. The statistical properties of CMB fluctuations depend on the primordial perturbations from which they arose, as well as on the subsequent evolution of the Universe as a whole. For cosmological models in which initial perturbations are of a Gaussian nature, the information carried by CMB anisotropies can be completely characterized by their angular power spectrum which depends on a few cosmological parameters. This makes the precise measurement of the CMB power spectrum a gold mine for understanding and describing the Universe throughout its history.\\
Pictures of the CMB maps delivered by the frequency channels of WMAP \citep{WMAP9_1} or Planck \citep{PR1_compsep} are contaminated by the astrophysical foreground emissions from our galaxy and extragalactic sources. The estimation of an accurate full-sky CMB map requires removing these emissions all over the sky \citep{Bouchet}. Computing a clean estimate of the CMB map on the Galactic center is particularly challenging. In addition, the instrumental noise hinders the estimation of the CMB map. In the low frequency regime (below ~100GHz, i.e. for WMAP or Planck LFI channels) the strongest contamination comes from the { Galactic} synchrotron and free-free emission \citep{Gold_WMAP7Templates}, with the highest contribution at large angular scales. Spinning dust \citep{PER_SDust} is an extra emission which spatially correlates with dust and dominates at low frequencies. At higher frequencies, the dust emissions \citep{PER_Dust} dominate whereas the synchrotron, free-free emissions and spinning dust are low.\\
The estimation of a clean foreground-free CMB map from the frequency channels is best performed by component separation techniques \citep{Leach08}. The CMB maps which have been made available by the Planck consortium have been obtained by using four different component separation methods \citep{PR1_compsep}:
%were used for Planck data processing, one template fitting method (Sevem), one method based 
%on a sky model (Commander-Ruler), and two Internal Linear Combination (NILC and SMICA) 
\begin{itemize}
\item{SEVEM:} SEVEM performs template fitting \citep{WSEVEM} in two distinct regions on both the 100 and 143 Ghz maps. Precisely, four templates are    
   derived from the difference of two channel maps (30-44), (44-70), (545-353) and (857-545).The final CMB { map} is then obtained by combining the two cleaned 100 and 143 Ghz maps.
\item{NILC:} is an ILC-based method which is performed in the wavelet domain \citep{NeedletILC}. The standard NILC approach is applied at each wavelet scale, and for different regions. Up to 20 regions were used at the finest scale. All Planck channels except the 30GHz are used.
\item{SMICA:} is a component separation method based on second-order statistics in the spherical harmonic domain \citep{ica:Del2003}. It includes a modeling of the foreground covariance matrix for $\ell < 1500$ and further performs ILC for $\ell > 1500$. All Planck channel are used.
\item{Commander-Ruler:} CR \citep{Commander} considers a sky modeling based on four components 
(CMB,  low-frequency emission, CO emission and thermal dust emission). Model parameters are derived at a 40 arc minute resolution, 
and the full resolution sky modeling is obtained by interpolating the parameters. Only channels whith frequencies ranging from $30$ to $353$GHz are used. 
\jbc{At the time  this study was done, the CR map was not yet available. This is the reason why it will not be discussed in this paper.}
% To our knowledge, the CR map has not been made available yet; it will not be discussed in this study.
\end{itemize}
The SEVEM map is a full-sky map. The SMICA and NILC maps are not full-sky where masked pixels (about $4 \%$ for SMICA) are inpainted or interpolated using a diffuse inpainting method. Furthermore, these maps are contaminated by the SZ effect (Sunyaev Zel'Dovich) which is problematic for CMB/SZ cross studies.\\
The quality of component separation methods highly { depends on} the number of degrees of freedom (d.o.f.'s) available to clean foreground contaminants. It is limited by the number of observed frequency channels. From this point, estimating a full-sky map with a clean { Galactic} center along with a low SZ contamination may sound like a dilemma: on the one hand further constraining the SZ effect requires freezing one d.o.f., on the other hand the estimation of a clean full-sky CMB map requires a copious number of d.o.f.'s to clean the complex emissivity variations of the { Galactic} center. Fortunately, additional d.o.f.'s can be {included} by combining {several} full-sky microwave surveys such as Planck PR1 and WMAP nine-year.

\subsection*{Contributions}
In this paper, we jointly process the WMAP nine-year and Planck PR1 data to recover a single CMB map. For this purpose, we make use of a recently introduced component separation method coined LGMCA \citep[Local Generalized Morphological Component Analysis,][]{2012arXiv1206.1773B}. Based on the concept of sparsity \citep{Starck2013}, the LGMCA radically departs from the methods used so far to estimate Planck CMB maps which all rely on second-order statistics. The combination of the Planck and WMAP data yields a CMB map with significant improvements:
\begin{itemize} 
\item full-sky map with no interpolated or inpainted pixels (Figure~\ref{fig:CMBmaps})
\item very clean estimation of the { Galactic region} with very low foreground-related artefacts (sections \ref{sec:QMAP}, \ref{sec:Galcenter})
\item low dust contamination for $\ell < 1000$ (section \ref{sec:Foregrounds})
\item virtually no SZ contamination (section \ref{sec:SZ})\\
\end{itemize}
Section \ref{sec:compsep} briefly describes the basics LGMCA method and the details of the joint processing of WMAP nine-year and Planck PR1. The map we derived from LGMCA is displayed and compared with the available Planck-only CMB maps in section~\ref{sec:results}.

\section{Sparsity and CMB Map Reconstruction}
\label{sec:compsep}
%The LGMCA algorithm\citep{2012arXiv1206.1773B} estimates both the components and the mixing matrix by maximizing the level of sparsity of each component; it seeks the sparsest sources possible in a wavelet basis. The observed sky is assumed to be a linear combination of all the components. Each component originates from completely different physical processes. The separation principle in LGMCA relies on the different spatial morphologies or structures of the various foregrounds, which translate into different sparsity patterns when transformed to a fixed wavelet dictionary. A mixture of these components decreases the level of sparsity.  LGMCA then estimates the components with maximal sparsity in the wavelet space. As emphasized in \citet{2012arXiv1206.1773B}, this is an efficient strategy to distinguish between physically different sources. 

The GMCA (Local-Generalized Morphological Component Analysis) method is based on 
blind source separation (BSS) \citep{2012arXiv1206.1773B}. In the framework of BSS, the observed sky is assumed to be a linear combination of { $m$ components} so that the $M$ frequency channels verify:
\begin{equation}
\forall i=1,\cdots,M; \, x_i = \sum_{j=1}^m\left( a_{ij} s_j + n_i\right)
\end{equation}
where $s_j$ stands for the $j$th component, $a_{ij}$ is a scalar that models for the contribution of the $j$-th component to channel $i$ and $n_i$ models the instrumental noise. This problem is more conveniently recast into the matrix formulation~:
\begin{equation}
{\bf X} = {\bf A S} + {\bf N}
\end{equation}
With the exception of the parameterized Bayesian method C-R (Commander-Ruler), currently available CMB maps which were derived from the Planck PR1 data are all based on the minimization of second-order statistics. In contrast, the GMCA method \citep{2012arXiv1206.1773B} relies on a radically different separation principle: sparsity. The fact that foreground components are sparse in the wavelet domain ({\it i.e.} a few wavelet coefficients are enough to represent most of the energy of the component) with different sparsity patterns {means} sparsity { acts} as a good separation criterion. {Taking} the data to the wavelet representation only alters the statistical distribution of the data coefficients without affecting its information content. A wavelet transform tends to grab the informative coherence between pixels while averaging the noise contributions, thus enhancing the structure in the data. This helps distinguishing different components that do not share the same sparse distribution in the wavelet domain. In addition, sparsity has the ability to be more sensitive to non-Gaussian processes, which has been shown to improve the foreground separation method. This is especially {true in} the {Galactic} center where the rapid variations of the emissivity of components like dust emissions or compact sources can be well measured by sparsity-based separation criteria.\\
Having $\boldsymbol{A}$ as the mixing matrix and $\boldsymbol{\Phi}$ as a wavelet transform, we assume that each source $s_{i}$ can be sparsely represented in ${ \bf \Phi}$; $s_{j}=\alpha_{j}\boldsymbol{\Phi}$, where $\mathbf{\alpha}$ is a $N_{s}\times T$ matrix whose rows are $\alpha_{j}$.
The multichannel noiseless data $\boldsymbol{Y}$ can be written as
\begin{equation}
\boldsymbol{Y}=\boldsymbol{A}\mathbf{\alpha}\boldsymbol{\Phi}\:.\label{eq:tensor1-1}
\end{equation}
The GMCA algorithm seeks an unmixing scheme which yields the sparsest sources $\boldsymbol{S}$. This is made formal by the following optimization problem (written in the Lagrangian
form)
\begin{equation}
\min\frac{1}{2}\left\Vert \boldsymbol{X}-\boldsymbol{A}\mathbf{\alpha}\boldsymbol{\Phi}\right\Vert _{F}^{2}+\lambda \left \| \mathbf{\alpha} \right \|_{p}^{p}\:,\end{equation}
where typically $p=0$ (or its relaxed convex version with $p=1$)
and ${\bf \left\Vert \boldsymbol{X}\right\Vert }_{\mathrm{F}}= \sqrt { \left(\textrm{trace}(\boldsymbol{X}^{T}\boldsymbol{X})\right)}$
is the Frobenius norm.\\ \\
The local-GMCA (LGMCA) algorithm \citep{2012arXiv1206.1773B} has been introduced as an extension { to} GMCA. Precisely, multi-frequency instruments generally provide observations with different resolutions. For example, the WMAP frequency channels have a resolution that ranges from $13.2$ arcmin for the $W$ band to $52.8$ arcmin for the K band. The Planck PR1 data have a resolution that ranges from $5$ arcmin at frequency $857$GHz to $33$ arcmin at frequency $30$GHz. The linear mixture model assumed so far in LGMCA no longer holds. This problem can be alleviated by degrading the frequency channels down to a common resolution before applying any component separation technique. The data are first decomposed in the wavelet domain. At each wavelet scale we only use the observations with invertible beams and then degrade the maps to a common resolution. This allows us to estimate a CMB map with a resolution of $5$ arcmin.\\ 
Furthermore, it is important to note that most foreground emissions ({\it e.g.} thermal dust, synchrotron, free-free, spinning dust) have electromagnetic spectra that are not spatially constant. As a consequence the mixing matrix $\bf A$ also varies across pixels, contrary to what assumed in GMCA. To deal with the spatial variation of the electromagnetic spectrum of some of the components, the LGMCA estimates the mixing matrices on patches at various wavelet scales with band-dependent size. An exhaustive description of LGMCA can be found in \citep{2012arXiv1206.1773B}. \\
The LGMCA algorithm has been implemented and evaluated on simulated Planck data in \citep{2012arXiv1206.1773B}.  It has also been applied to the WMAP nine-year data \citep{WMAP9_LGMCA}.

\subsection{LGMCA parameters for the joint processing of  WMAP and Planck data}
\begin{figure}[htb]
\centerline{\includegraphics [scale=0.2]{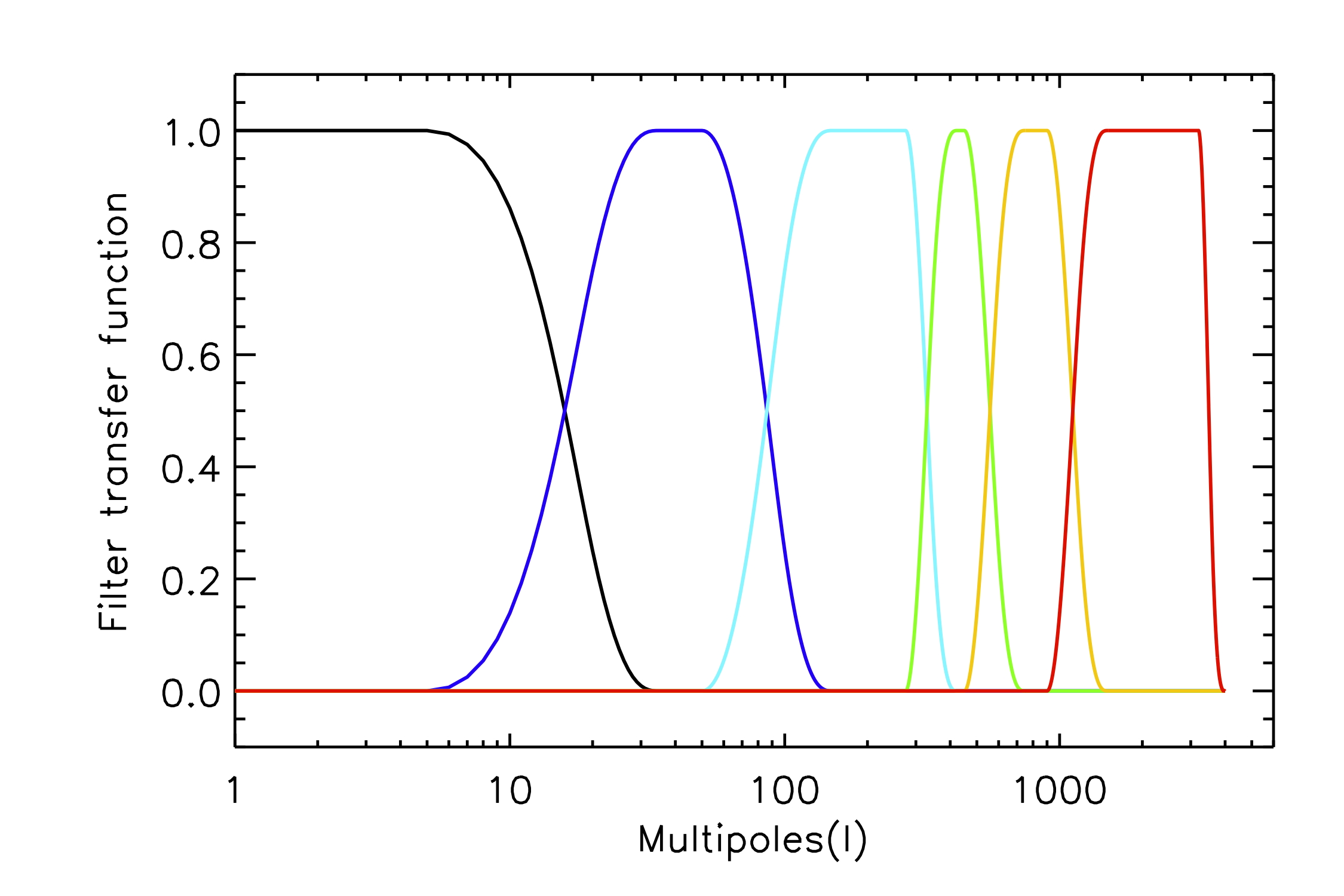}}
\caption{Transfer functions of the $6$ wavelet bands used to estimation the CMB with LGMCA.}
\label{fig:bands}
\end{figure}

\begin{figure}[htb]
\centerline{\includegraphics [scale=0.5]{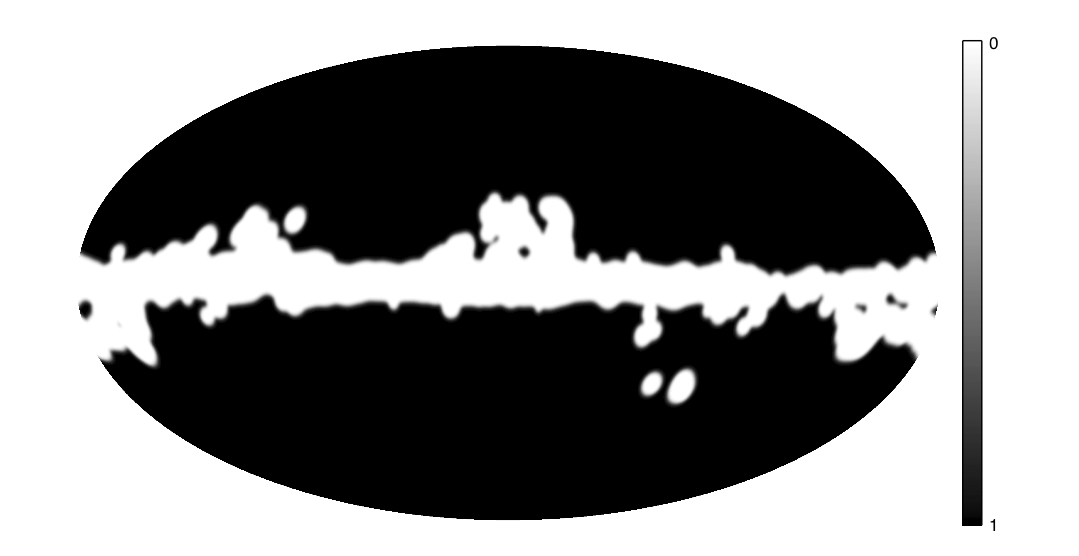}}
\caption{Mask used for the specific processing of the Galactic center. Precisely, the IRIS map is used as an extra observation in the Galactic part of the mask. The sky coverage is about $\mbox{\it fsky} = 82\%$.}
\label{fig_mask_gal}
\end{figure}

The LGMCA mixing matrices are estimated from a set of input channels at a given resolution on a patch of data at a given wavelet scale. The parameters used by LGMCA to process {jointly} the WMAP nine-year and Planck PR1 data are described in Table~\ref{tab_resol_planck}. Figure~\ref{fig:bands} displays the filters in spherical harmonics defining the wavelet bands at which the derived weights (by inverting these mixing matrices) were applied.\\
Contrary to the Planck-based CMB maps, most estimates of the CMB maps which were computed from the WMAP data made use of ancillary data to improve the cleaning of {Galactic} foreground emissions (see \citep{WMAP9_1,2012MNRAS.419.1163B}). In \citep{WMAP9_LGMCA}, the use of a dust template helped improving the quality of the estimated CMB map. When it turns to the analysis of Planck data, increasing the number of observations by using ancillary observations can be fruitful as well. This is especially true in the Galactic center where the linear mixture model used so far in component separation methods is more akin to fail. For that purpose, the IRIS map \citep{IRIS05} is added as an extra observation in an area defined by the mask in Figure~\ref{fig_mask_gal}; this allows to increase the number of d.o.f.'s which greatly helps cleaning the CMB in the Galactic center. It is however important to keep in mind that the power spectrum of the proposed CMB map will be computed from a sky area where the CMB map is evaluated without the IRIS map.

%The parameters used to process the WMAP data are described in Table ~\ref{tab_resol_planck}. LGMCA is then applied on $6$ different wavelet bands; these bands are entirely described by filters in spherical harmonics, shown in Figure~\ref{fig:bands}.
% LGMCA is applied on $6$ different wavelet bands; these bands are entirely described by filters in spherical harmonics, shown in Figure~\ref{fig:bands}.
% In each band, only a subset of the data is used and downgraded to a common resolution defined as the lowest resolution of the maps within the selected subset. % [JLS modif] Did not want to speak about the wavelet decomposition. So I modified also the following table too
%In each band, mixing matrices are estimated in the wavelet domain with a band-dependent number of scales (WT scales).
%  As described in \citep{2012arXiv1206.1773B}, the mixing matrices are estimated on square patches the size of which also depends on the band. The parameters used to process the WMAP data are described in Table ~\ref{tab_resol_planck}.

\begin{table}
\begin{center}
\vspace{0.1in}
\begin{tabular}{|c|c|c|c|c|}
\hline
Band &  WMAP Obs. & Planck Obs. & Patch size  & Res. \\
\hline
\hline
I & All  & All  & None & 60  \\
 \hline
II & Q,V,W  & All & 64 & 33 \\
 \hline
III & V,W bands &  $44$ to $857$GHz  & 64  & 24 \\
 \hline
IV & W bands & $70$ to $857$GHz  & 32  & 14 \\
 \hline
V & No & $100$ to $857$GHz  & 16 & 10 \\
  \hline
VI & No & $143$ to $857$GHz & 8  & 5 \\
\hline
\end{tabular}
\vspace{0.1in}
\end{center}
\caption{Parameters of LGMCA to process the WMAP nine-year and PR1 data. For each band, the second ({\it resp.} third) column gives the subset of WMAP ({\it resp.} Planck) data used to analyze the data, the fourth column provides the size of the square patches at the level of which the analysis is made, the last column gives the common resolution of the data in arcmin.}
\label{tab_resol_planck}
\end{table}

\subsection{Compact sources and Galactic center post-processing}

\begin{figure}[htb]
\centerline{\includegraphics [scale=0.5]{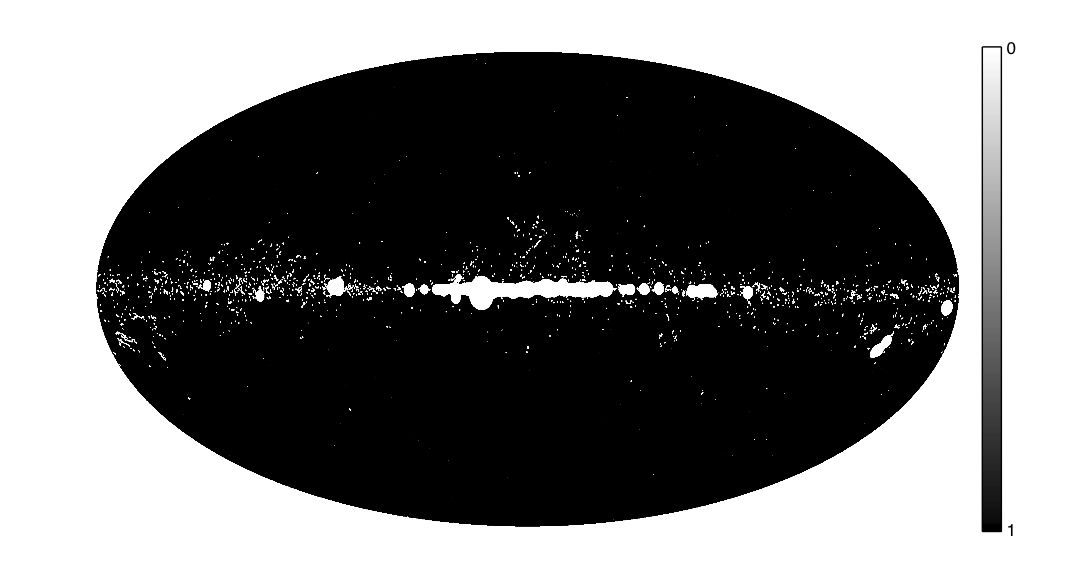}}
\caption{Mask used for the post-processing of the compact sources. The sky coverage is equal to $\mbox{\it fsky} = 97\%$.}
\label{fig:mca_mask}
\end{figure}

Combining WMAP and Planck is not sufficient to properly clean for compact structures like compact sources especially on the Galactic center where they can be found in large number. The performances of the linear mixture model used so far in LGMCA turns out to be quite limited to extract these types of contaminants as it would require far more d.o.f.'s. This limitation can be alleviated by switching to { a} non-linear estimator to clean for the compact sources that still contaminate the LGMCA CMB map estimate. Separating the compact sources from the raw CMB map estimate can be recast as a single-observation component separation. This problem is tackled by the MCA \citep[morphological component analysis,][]{inpainting:abrial06}. In this framework, the raw estimate of the CMB map $x$ is assumed to be the linear combination of the compact sources signal $x_p$ and the clean CMB map $x_c$: $x = x_p + x_c$. From \citep{inpainting:abrial06}, emphasizing on the morphological differences of the compact sources and the CMB map allows for an accurate separation of both signals. Precisely, the compact sources signal is sparse in the wavelet domain $\bf \Phi$ and restricted to compact regions while the CMB map is homogeneous across the sky and sparsely distributed in the Spherical harmonics domain $\bf F$. The MCA estimate of the compact sources signal $x_p$ and the CMB map $x_c$ is given by the solution of the following optimization problem:
\begin{equation}
\label{eq:mca}
\min_{x_c, x_p} \|x_c {\bf F}^T\|_1 + \| x_p {\bf \Phi}^T\|_1 \mbox{ s.t. } x = x_c + x_p ; x_p[\Omega] = 0
\end{equation}
where the constraint $x_p[\Omega] = 0$ enforces the compact source signal to be zero outside of a prescribed region complementary to $\Omega$.
%non vanishing in a prescribed sky region which is complementary to $\Omega$. 
In practice the set $\Omega$ and its complement are defined by the compact sources mask and a very restricted region about the Galactic center as displayed in Figure~\ref{fig:mca_mask}. Interestingly, even if the solution to the problem in Equation~\ref{eq:mca} provides solutions which depend non-linearly on the data $x$, it provides a linear decomposition via the constraint $x = x_c + x_p$. It is also important to notice that pixels of the clean CMB map $x_c$ in the sky region $\Omega$ remain unaltered by this mechanism.

\subsection{Map and Power Spectrum Estimation}
\begin{figure}[htb]
\hbox{
\centerline{\includegraphics [scale=0.5]{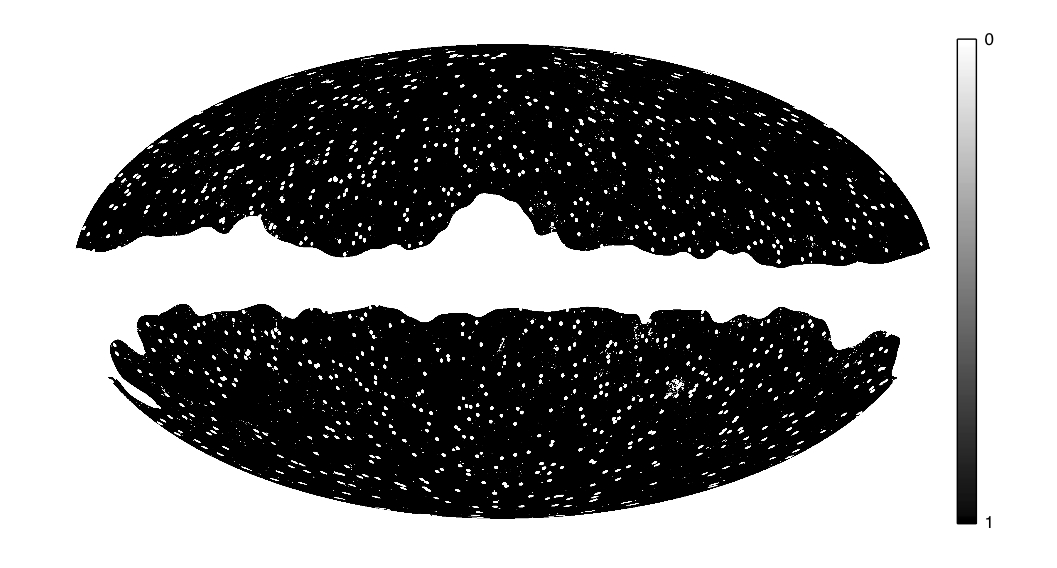}}
}
\caption{Mask used for power spectrum estimation. The sky coverage is equal to $\mbox{\it fsky} = 76\%$.}
\label{fig_maskps}
\end{figure}

\begin{figure*}[htb]
\centerline{
\hbox{
\includegraphics [scale=0.2]{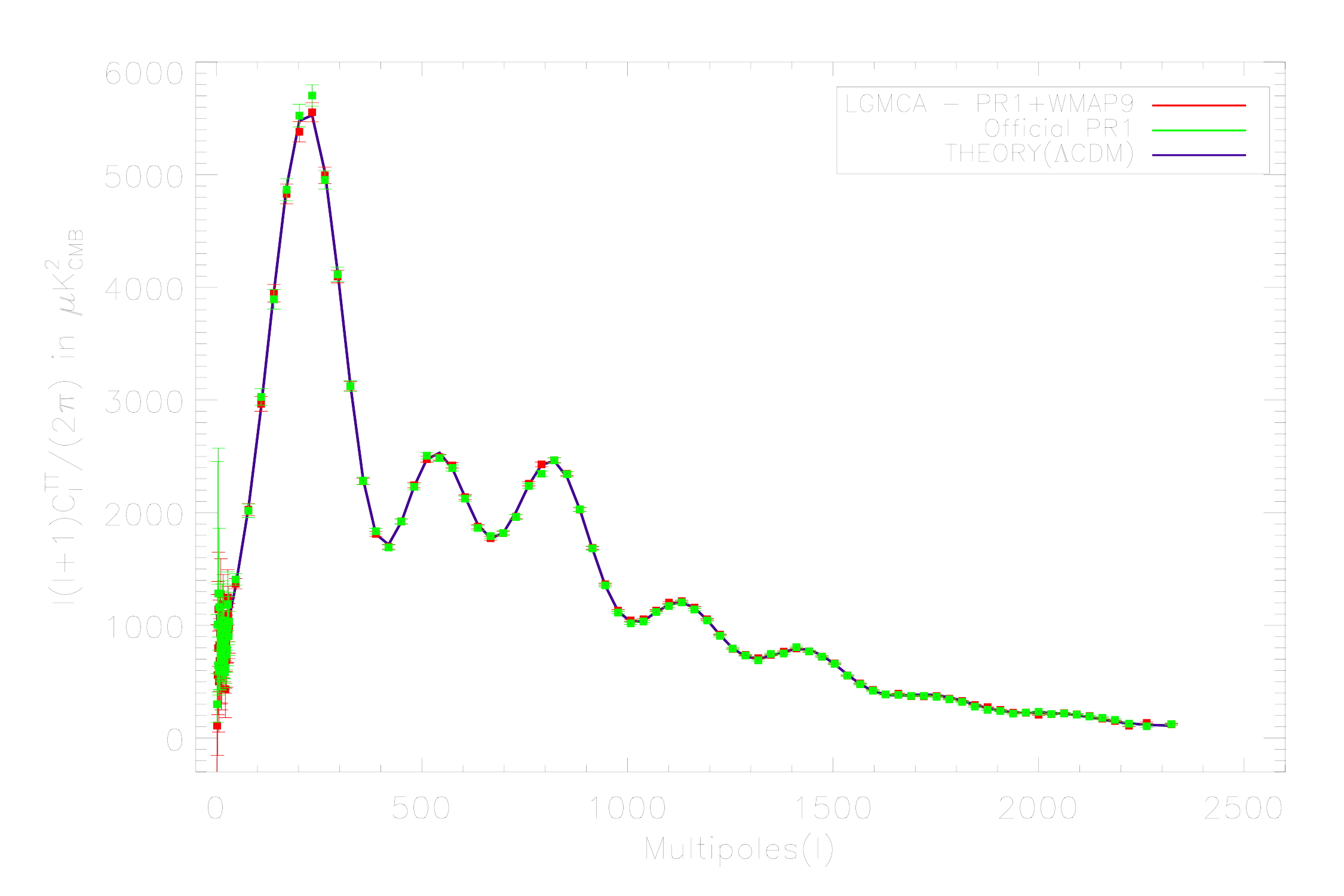}
\includegraphics [scale=0.2]{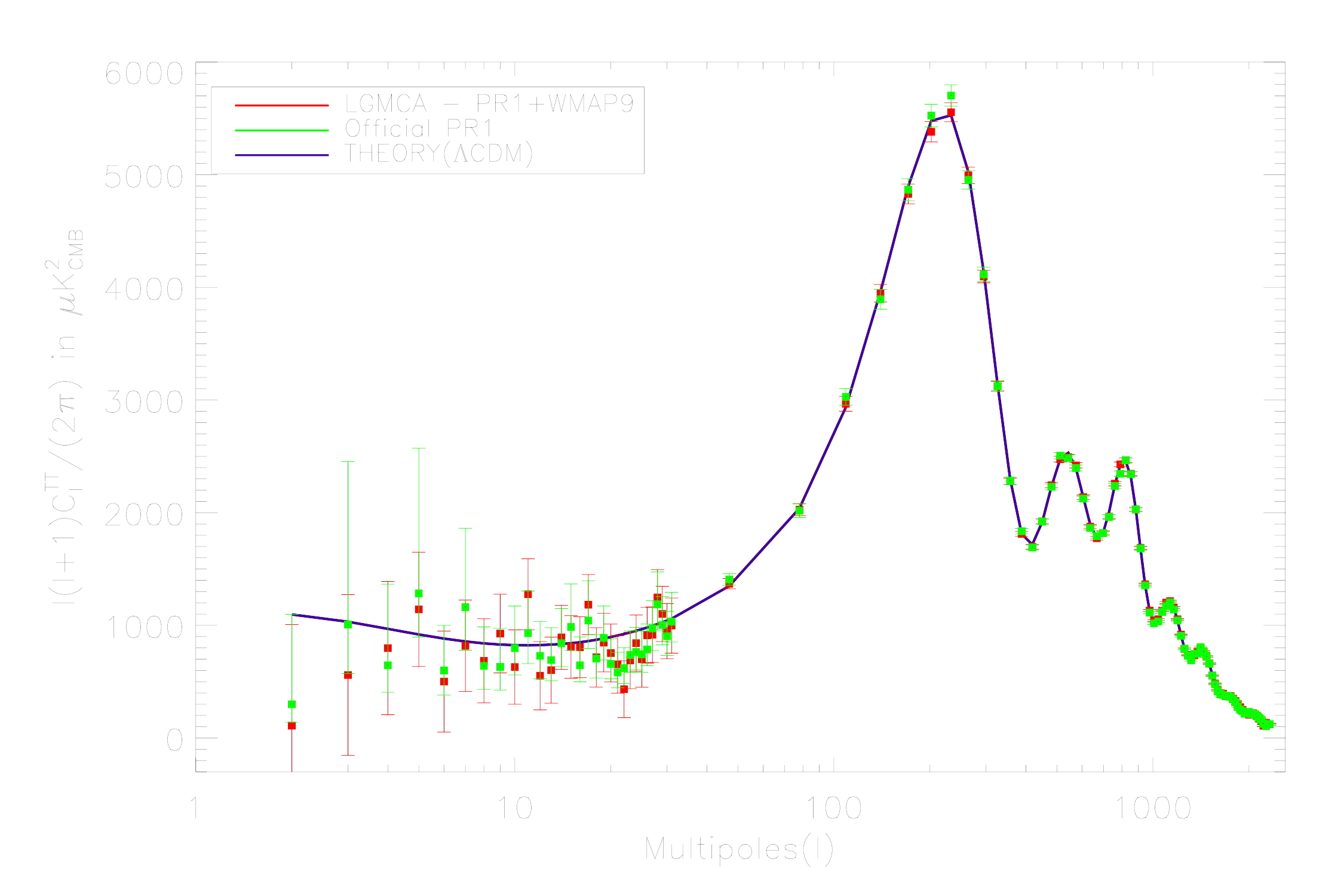}
}}
\caption{Left, estimated power spectrum of the WPR1 LGMCA map (red) and official PR1 power spectrum (green). The solid black line is the Planck-only best-fit $C_\ell$ provided by the Planck consortium. Right, power spectrum in logarithmic scale. Error bars are set to $1\, \sigma$.}
\label{fig_ps}
\end{figure*}

\begin{figure*}[htb]
\hbox{
\centerline{
\includegraphics [scale=0.2]{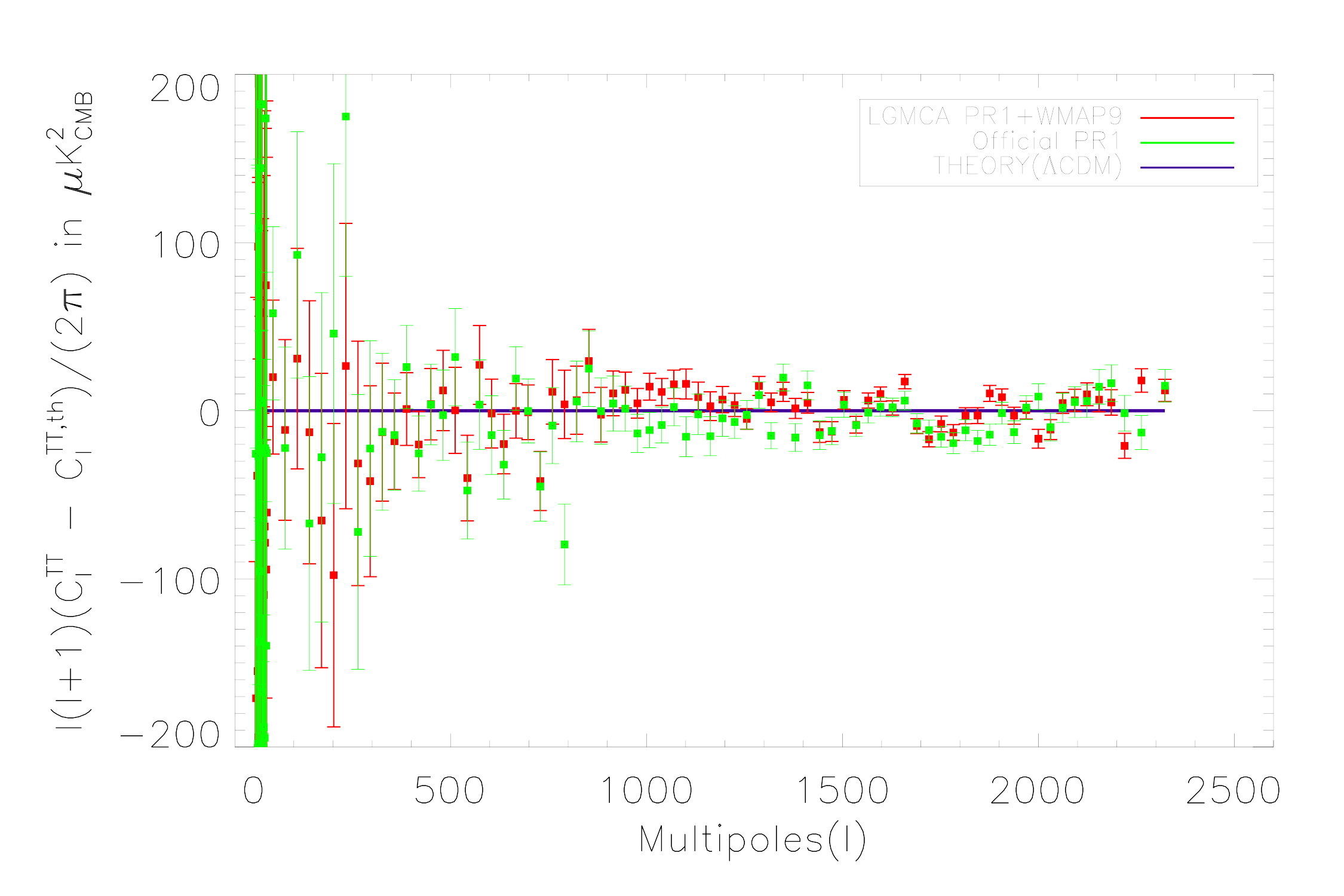}
\includegraphics [scale=0.2]{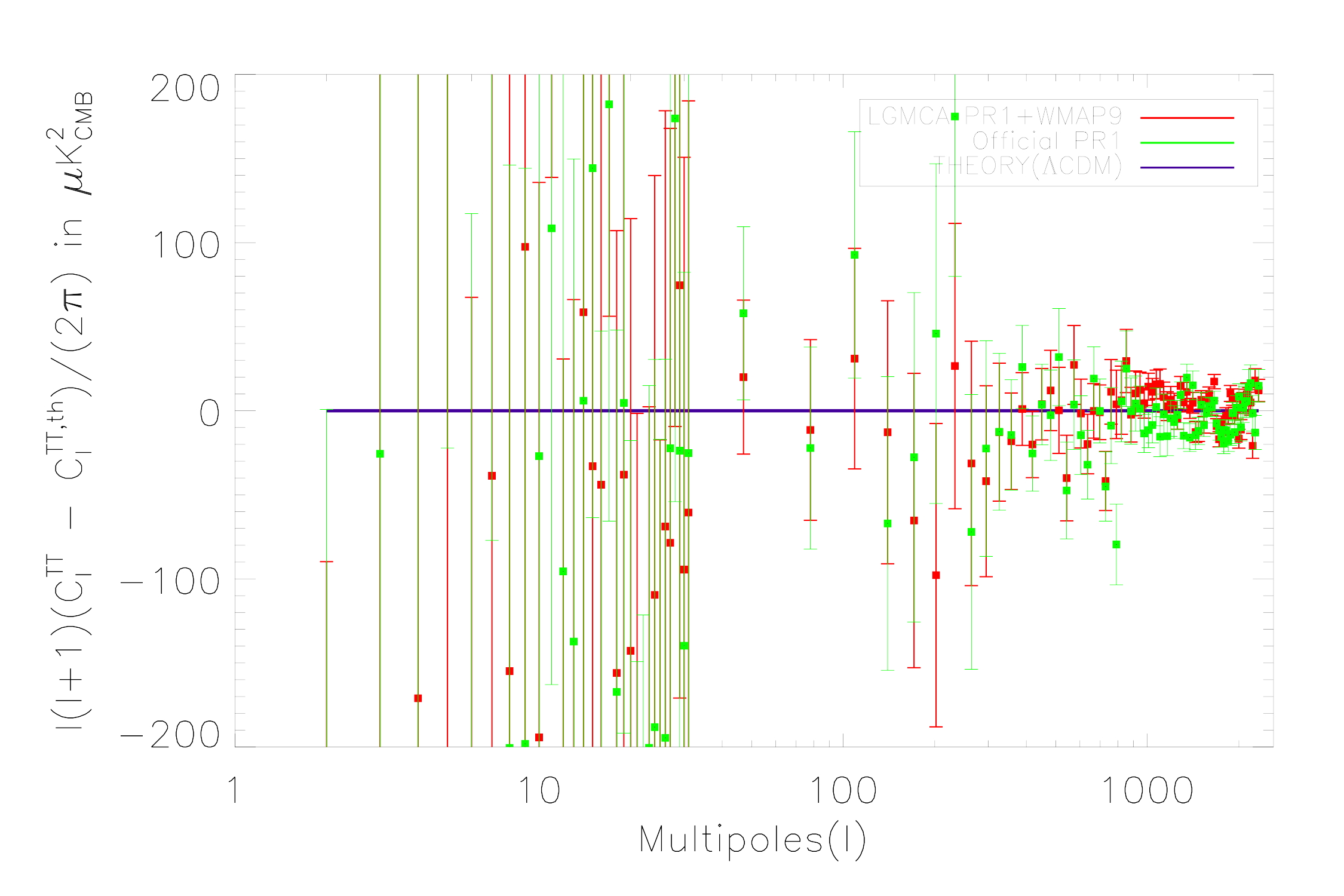}
}}
\caption{Left, difference between the power spectrum estimated from the WPR1 LGMCA map (red) ({\it resp.} official PR1 power spectrum (green)) and the Planck-only best-fit $C_\ell$ provided by the Planck consortium. Right, difference between the estimated and theoretical power spectra in logarithmic scale. Error { bars} are set to $1\, \sigma$.}
\label{fig_resi_ps}
\end{figure*}

Following \citep{2012arXiv1206.1773B}, the LGMCA is applied to the five WMAP maps and the nine averages of  Planck PR1 half ring maps so as to estimate the set of mixing matrices. The pseudo inverse of these mixing matrices are then applied to the same WMAP and Planck data to estimate the CMB map. Noise maps are generally derived by applying the pseudo-inverse of the mixing matrices to noise realizations of the data. In the case of WMAP, random noise realizations are computed using the noise covariance matrices which have been provided by the WMAP consortium. For Planck, half differences of half ring maps provide a good proxy for a single data noise realization.\\
Next in this article, the CMB power spectrum is estimated by computing the cross-correlation between the two half-ring maps. In contrast to the CMB signal, noise decorrelates between half rings; the noise bias then vanishes when cross-correlating half-ring maps. Such cross-correlation therefore provides an estimate of the CMB power spectrum which turns to be free of any bias from the noise. In the case of WMAP, such virtual half rings maps can be obtained by calculating the difference and sum of the WMAP data and a single data noise realization.\\
The power spectrum is evaluated from a sky coverage of $76\%$; the corresponding mask is composed of a point sources and { Galactic} mask { chosen} from the Planck consortium masks.The mask we used for power spectrum estimation is displayed in Figure~\ref{fig_maskps}. Prior to computing the cross-correlation between the half-ring maps, the maps are deconvolved to infinite resolution up to $\ell = 3200$. Changing the resolution of maps is very likely to create artifacts especially at the vicinity of remaining point sources. To alleviate this problem, the masked map are first inpainted prior to deconvolution. It is important to point out that this stage does not alter the estimation of the power spectrum as it is eventually evaluated from the $76\%$ of the sky which are kept unchanged through the inpainting step. Correcting for the effect of the mask is made by applying the MASTER mask deconvolution technique \citep{2002ApJ...567....2H}. \\ \\

The CMB power spectrum is biased by the contamination of the unresolved point sources, especially at high $\ell > 1500$. \jbc{In \citep{PR1_PS}, unresolved point sources are modeled as a contamination with constant power spectrum in each frequency channels. After component separation, the estimated CMB map is computed as a linear combination of the frequency channels. Furthermore, the CMB power spectrum is estimated from regions of the sky (described in Figure \ref{fig_maskps}) where LGMCA parameters are very likely constant in each wavelet band. As a consequence, the power spectrum of the unresolved point sources will be constant in each of the $6$ wavelet bands. These scalar parameters are denoted by $\{\mathcal{A}_\ell^s\}_{s=1,\cdots,6}$ in sequel.\\
Following \citep{PR1_PS}, an accurate correction of the point sources contribution would require estimating the parameters $\{\mathcal{A}_\ell^s\}_{s=1,\cdots,6}$ together with the cosmological parameters by maximizing their joint likelihood; this is beyond the scope of this paper. As first-order correction, these point sources parameters are estimated by minimizing the least-square minimization of the error $C_\ell - C_\ell^{th}$ where $C_\ell$ stands for the estimated power spectrum and $C_\ell^{th}$ for the best-fit Planck power spectrum. For instance, the resulting point sources parameter in the latest wavelet band ($s=6$) takes the value $\ell (\ell + 1)/2\pi \mathcal{A}_\ell^{s=6} = 174 \, \mu K^2$ for $\ell = 3000$ (following the convention defined in \citep{PR1_compsep}). This value is of the same order of magnitude as those obtained for other component separation methods in the Planck component separation paper (see the parameter $A_{ps}$ in Figure 11 of \citep{PR1_compsep}).}\\ 
The estimated CMB power spectrum as well as the official Planck power spectrum are displayed in Figure~\ref{fig_ps}. The error bars of our estimate of the power spectrum account for the cosmic variance and noise only. Slight differences between the Planck best fit and Planck+WMAP9 power spectra can be seen at very low $\ell$; this can be seen with more precision in Figure~\ref{fig_resi_ps}.\\
In the next, the compatibility of the estimated power spectrum with the official Planck best-fit is evaluated. To that end, a standard $\chi^2$-based goodness-of-fit procedure and a error tail statistics evaluation are carried out. In the sequel, the error between the estimated and Planck best-fit power spectra is defined as :
$$
\mathcal{E}_\ell = \frac{C_\ell - C_\ell^{th}}{\sqrt{V_\ell}}
$$
where $V_\ell$ denotes the variance of the estimated power spectrum. In case of compatibility, the error $\mathcal{E}_\ell$ should be distributed according to a standard normal distribution with mean zero and variance one. This can be tested by computing the error $\chi^2$ and the p-value of the resulting value. This test has been carried out on various ranges of multipoles $[2,\ell_{\max}]$ (the monopole and dipole are not taken into account) with $\ell_{\max} = 1500, 2000, 2500$. The results are displayed in table~\ref{tab_chi2}. For $\ell > 1500$, the $\chi^2$ test do not indicate a good match between the estimated and best-fit Planck power spectra; the corresponding p-values are larger than $0.1$.\\
The evaluation of the tail statistics of the error $\mathcal{E}_\ell$ provides a complementary compatibility test. The third ({\it resp.} fourth) row of table~\ref{tab_chi2} gives the number of samples of $\mathcal{E}_\ell$ with amplitudes higher than $3 \sigma$ ({\it resp.} $4 \sigma$). The probability of the observed number of samples follows a binomial distribution with known parameter; this quantity is also provided. For $\ell_{\max} \geq 1500$, the observed values are clearly not compatible with the expected theoretical values.\\ \\
This study has further been carried out on the binned power spectrum -- displayed in Figures~\ref{fig_ps} and \ref{fig_resi_ps}. The compatibility check results are featured in Table~\ref{tab_chi2_binned}. Despite the good $\chi^2$ values for all ranges of $\ell_{\max}$, the error $\mathcal{E}_\ell$ exhibits more extreme values than predicted by theory. This indicates that the estimated and best-fit Planck spectra are not compatible.\\
The discrepancy between the estimated power spectrum and the Planck best-fit may have different origins: i) inaccurate error estimation: the uncertainty of the power spectrum should account for the full covariance matrix, ii) inaccurate correction for unresolved point sources and/or CIB, iii) beam transfer function errors at larger $\ell$ should be taken into account to only name a few. These different sources of uncertainties will very likely increase the error bars at medium and small scales. Moreover, the error bars of the official power spectrum (displayed in green in Figure~\ref{fig_ps}) that includes also foreground and beam uncertainties are significantly larger than the error bars estimating only in this work cosmic variance and noise. If the estimated error bars are further substituted with the error bars of the official power spectrum, the $\chi^2$ test as well as the tail statistics from the binned power spectrum do not indicate anymore any incompatibility -- see the values in parenthesis in Table~\ref{tab_chi2_binned}. This suggests that the estimated error bars are probably too optimistic at small scales and should further be updated to account for foreground residuals -- namely point sources and CIB -- and instrumental uncertainties. 

%\begin{table}
%\begin{center}
%\vspace{0.1in}
%\begin{tabular}{|c|c|c|c|}
%\hline
%Band &  1500 & 2000 & 2500  \\
%\hline
%\hline
%$\chi^2$ p-value & $0.29$  & $0.22$  & $0.04$  \\
% \hline
%$\#(|\mathcal{E}_\ell| > 3)$  & $5$  & $17$ & $40$ \\
%Theoretical   & $4 \pm 3.9$  & $5.4 \pm 4.5$ & $6.7 \pm 5.1$ \\
% \hline
%$\#(|\mathcal{E}_\ell| > 4)$ &  $0$ & $3$  & $7$ \\
%Theoretical & $0.09 \pm 0.6$  & $0.12 \pm 0.69$ & $0.16 \pm 0.78$ \\
%\hline
%\end{tabular}
%\vspace{0.1in}
%\end{center}
%\caption{Compatibility check of the estimated power spectrum with the best-fit Planck PR1 from the single $\ell$ (unbinned) power spectrum. {\bfseries Second row :} the p-value of the $\chi^2$ for different ranges of multipoles. {\bfseries Third row :} ({\it resp.} fourth) row shows the number of samples of the normalized error $\mathcal{E}_\ell$ with amplitudes higher that $3 \sigma$ ({\it resp.} higher than $4 \sigma$) with respect to the theoretical value (binomial distribution) with a $95 \%$ confidence interval.}
%\label{tab_chi2}
%\end{table}

\begin{table}
\begin{center}
\vspace{0.1in}
\begin{tabular}{|c|c|c|c|}
\hline
Band &  1500 & 2000 & 2500  \\
\hline
\hline
$\chi^2$ p-value & $0.43$  & $0.1$  & $0.02$  \\
 \hline
$\#(|\mathcal{E}_\ell| > 3)$  & $8$  & $21$ & $44$ \\
Theoretical   & $4$  & $5.4$ & $6.7$ \\
Probability of event   & $0.03$  & $1.9\mbox{\sc{e}-}7$ & $1.0\mbox{\sc{e}-}21$ \\
 \hline
$\#(|\mathcal{E}_\ell| > 4)$ &  $0$ & $4$  & $8$ \\
Theoretical & $0.09$  & $0.12$ & $0.16$ \\
Probability of event   & $0.9$  & $9.4\mbox{\sc{e}-}6$ & $8.2\mbox{\sc{e}-}12$ \\
\hline
\end{tabular}
\vspace{0.1in}
\end{center}
\caption{Compatibility check of the estimated power spectrum with the best-fit Planck PR1 from the single $\ell$ (unbinned) power spectrum. {\bfseries Second row :} the p-value of the $\chi^2$ for different ranges of multipoles. {\bfseries Third row :} ({\it resp.} {\bf fourth}) row shows the number of samples of the normalized error $\mathcal{E}_\ell$ with amplitudes higher that $3 \sigma$ ({\it resp.} higher than $4 \sigma$) with respect to the theoretical value (binomial distribution). The probability of the observed number of extreme values, assuming that the error follows a standard Gaussian distribution, is also provided.}
\label{tab_chi2}
\end{table}

%\begin{table}
%\begin{center}
%\vspace{0.1in}
%\begin{tabular}{|c|c|c|c|}
%\hline
%Band &  1500 & 2000 & 2500  \\
%\hline
%\hline
%$\chi^2$ p-value & $0.06 \, (0.71)$ & $0.04 \, (0.65)$ & $0.07 \, (0.77)$ \\
% \hline
%$\#(|\mathcal{E}_\ell| > 3)$  & $0 \, (0)$  & $2 \, (1)$ & $3 \, (1)$ \\
%Theoretical   & $0.2 \pm 0.9$  & $0.25 \pm 0.98$ & $0.28 \pm 1.03$ \\
% \hline
%$\#(|\mathcal{E}_\ell| > 4)$ &  $0 \, (0)$ & $1 \, (0)$  & $1 \, (0)$ \\
%Theoretical & $0.0048 \pm 0.14$  & $0.0058 \pm 0.15$ & $0.0065 \pm 0.16$ \\
%\hline
%\end{tabular}
%\vspace{0.1in}
%\end{center}
%\caption{Compatibility check of the estimated power spectrum with the best-fit Planck PR1 from the binned power spectrum. {\bfseries Second row :} the p-value of the $\chi^2$ for different ranges of multipoles. {\bfseries Third row :} ({\it resp.} fourth) row shows the number of samples of the normalized error $\mathcal{E}_\ell$ with amplitudes higher that $3 \sigma$ ({\it resp.} higher than $4 \sigma$) with respect to the theoretical value (binomial distribution) with a $95 \%$ confidence interval. {\bf Values in parenthesis :} these values are obtained by accounting for the error bars of the official power spectrum instead of the error bars derived from noise only.}
%\label{tab_chi2_binned}
%\end{table}

\begin{table}
\begin{center}
\vspace{0.1in}
\begin{tabular}{|c|c|c|c|}
\hline
Band &  1500 & 2000 & 2500  \\
\hline
\hline
$\chi^2$ p-value & $0.56 \, (0.81)$ & $0.42 \, (0.84)$ & $0.57 \, (0.74)$ \\
 \hline
$\#(|\mathcal{E}_\ell| > 3)$  & $0 \, (0)$  & $2 \, (1)$ & $2 \, (1)$ \\
Theoretical   & $0.2$  & $0.25$ & $0.28$ \\
Probability of event   & $0.81 \, (0.81)$  & $2.5\mbox{\sc{e}-}2 \, (0.2)$ & $2.9\mbox{\sc{e}-}3 \, (0.21)$ \\
 \hline
$\#(|\mathcal{E}_\ell| > 4)$ &  $0 \, (0)$ & $1 \, (0)$  & $1 \, (0)$ \\
Theoretical & $0.0048$  & $0.0058$ & $0.0065$ \\
Probability of event   & $0.91 \, (0.91)$  & $0.11 \, (0.88)$ & $0.13 \, (0.85)$ \\
\hline
\end{tabular}
\vspace{0.1in}
\end{center}
\caption{Compatibility check of the estimated power spectrum with the best-fit Planck PR1 from the binned power spectrum. {\bfseries Second row :} the p-value of the $\chi^2$ for different ranges of multipoles. {\bfseries Third row :} ({\it resp.} {\bf fourth}) row shows the number of samples of the normalized error $\mathcal{E}_\ell$ with amplitudes higher that $3 \sigma$ ({\it resp.} higher than $4 \sigma$) with respect to the theoretical value (binomial distribution). The probability of the observed number of extreme values, assuming that the error follows a standard Gaussian distribution, is also provided. {\bf Values in parenthesis :} these values are obtained by accounting for the error bars of the official power spectrum instead of the error bars derived from noise only.}
\label{tab_chi2_binned}
\end{table}

In the remaining of this paper the CMB map estimated from Planck and WMAP9 will be denoted by WPR1 LGMCA.

%%%%%%%%%%%%%%%%%%%%%%%%%%%%%%%%%%%%%%%%%%%%%%%%%%%%%%%%%%%%%%%%%%

\section{CMB maps evaluation}
\label{sec:results}

\begin{figure*}[htb]
\centerline{
\hbox{
\includegraphics [scale=0.25]{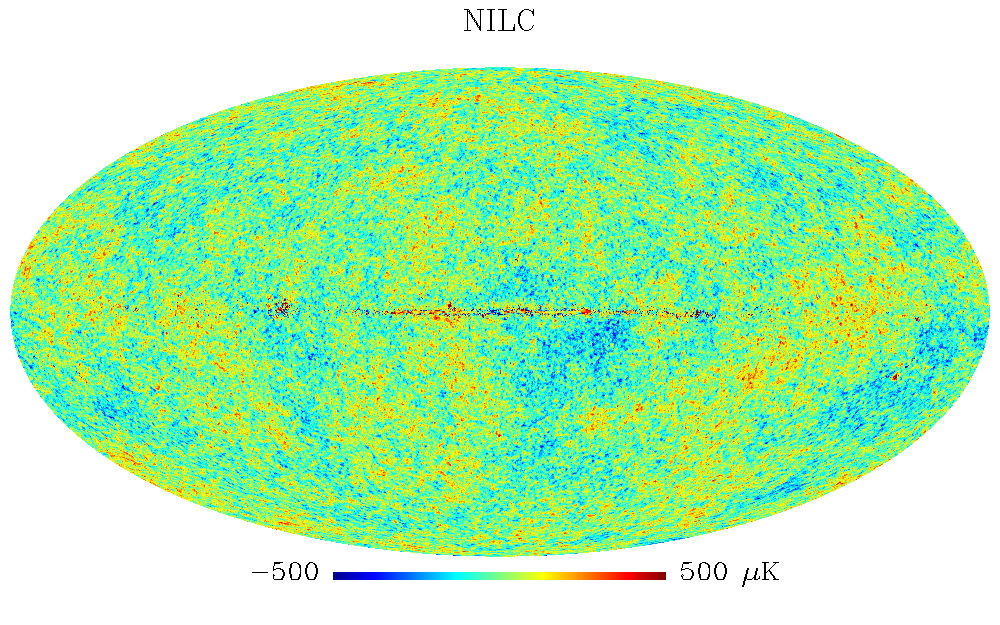}
\includegraphics [scale=0.25]{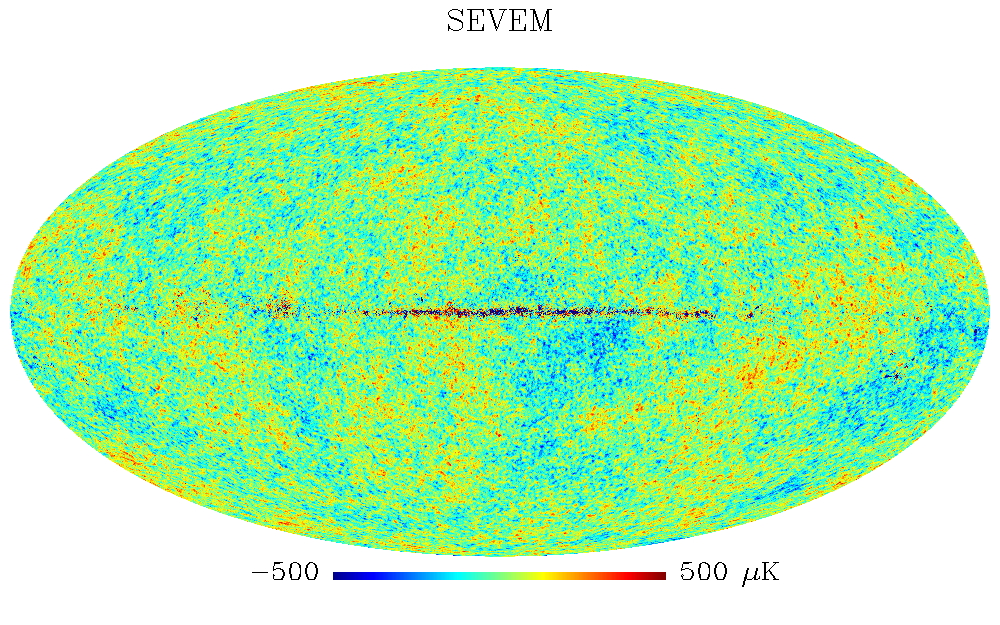}
}}
\centerline{
\hbox{
\includegraphics [scale=0.25]{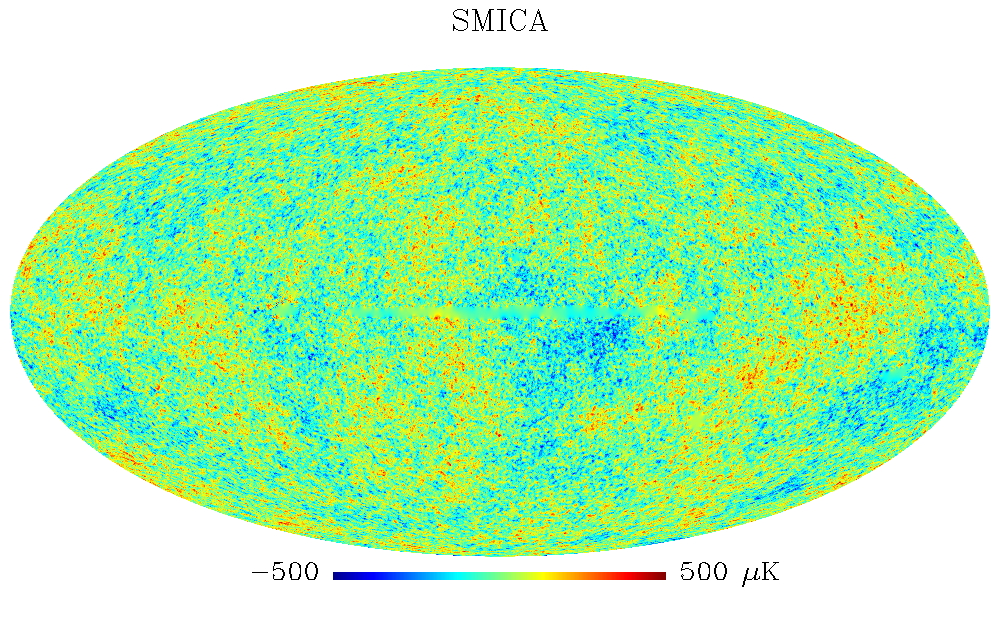}
\includegraphics [scale=0.25]{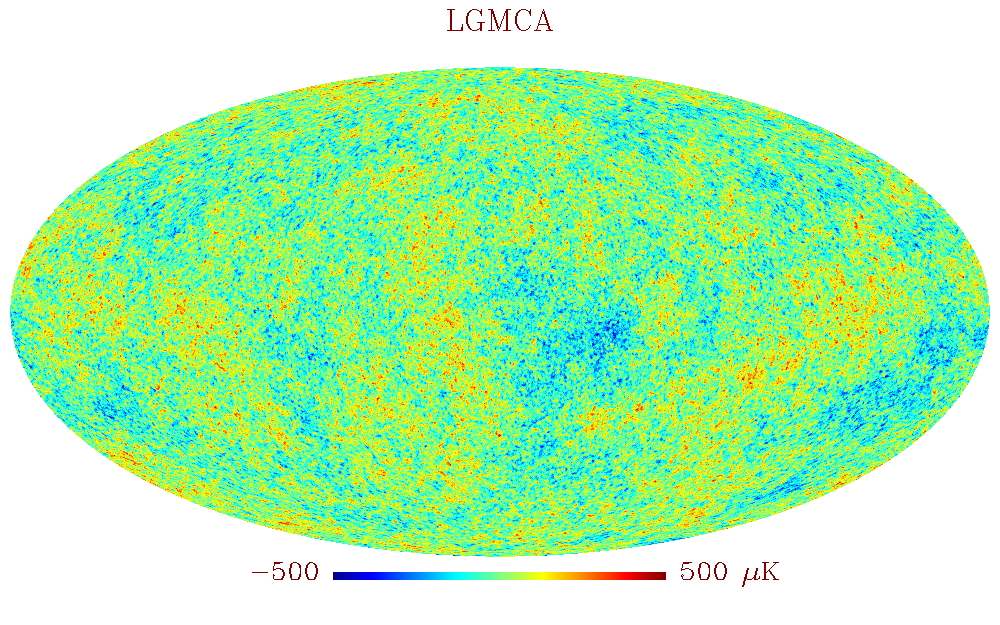}
}
}
\caption{Top, PR1  NILC and SEVEM CMB maps. Bottom, PR1  SMICA and WPR1 LGMCA CMB maps.}
\label{fig:CMBmaps}
\end{figure*}

\begin{figure*}[htb]
\centerline{
\hbox{
\includegraphics [scale=0.3]{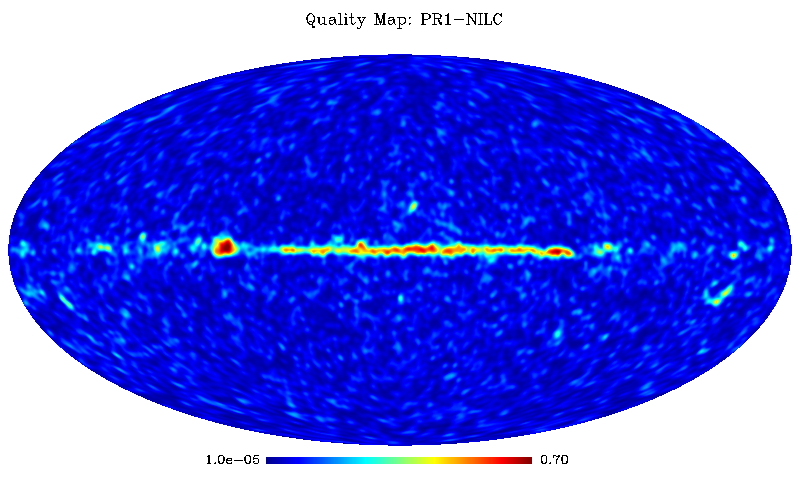}
\includegraphics [scale=0.3]{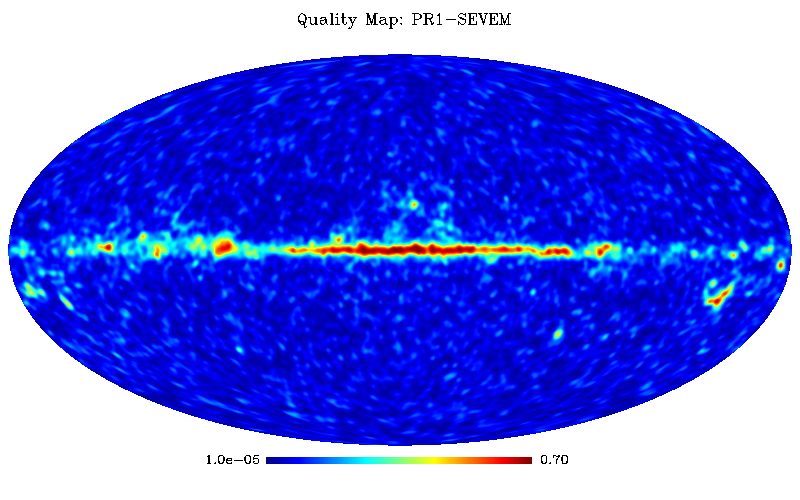}
}}
\centerline{
\hbox{
\includegraphics [scale=0.3]{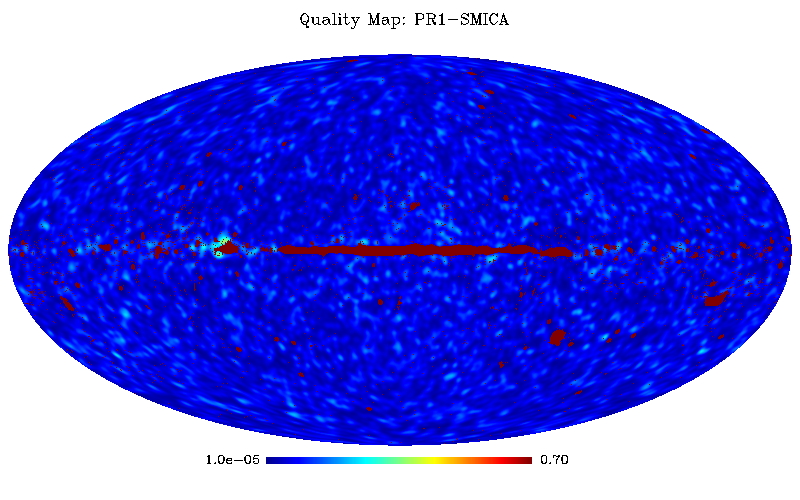}
\includegraphics [scale=0.3]{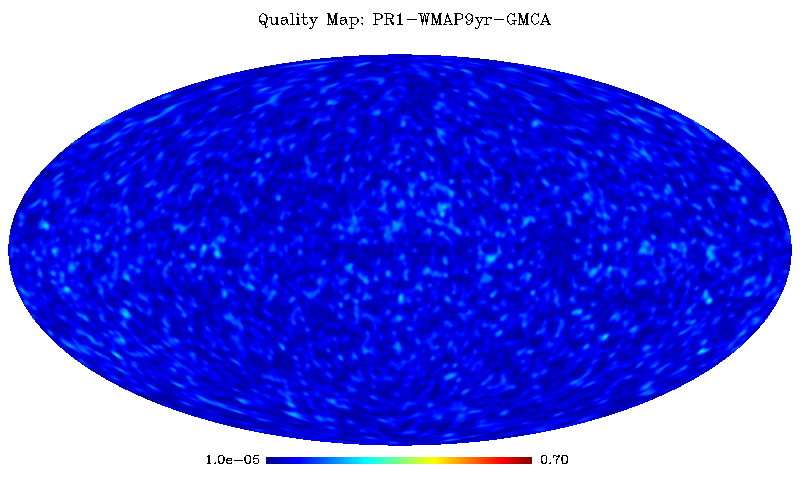}
}
}
\caption{Top, PR1  NILC and SEVEM quality maps. Bottom, PR1  SMICA and WPR1 LGMCA quality maps.}
\label{fig_qualmap}
\end{figure*}

This study aims to analyze the joint processing of the WMAP nine-year and Planck PR1 data so as to produce a clean and accurate estimate of the CMB on the entire sky. The larger number of d.o.f.'s that the combination of the WMAP and Planck data affords should allow for a better cleaning of the Galactic center as well as prevent the CMB map estimate from SZ residuals. The forthcoming comparisons will therefore precisely focus on evaluating deviations from the expected characteristics of the CMB map through the assessment of various measures of contamination signatures.

\subsection{\label{sec:QMAP}Measuring excess of power with the Quality Map}
Estimating the quality of an estimated CMB map $s$ from real data without any strong assumption about the expected map is challenging. In this section, we will assume that the $\lambda$-CDM best fit $C_\ell$ provides a power spectrum 
%of map 
that gives a good approximation to the expected power of the CMB per frequency.  This allows to compare the local deviation around each pixel $k$ in $s$ to this expected power and check whether it is compatible with the expected noise level that the best-fit $C_\ell$ indicates.\\
This method can be refined by performing this test in the wavelet space rather than in direct space. Hence, given a wavelet function $\psi$ and a given wavelet scale $j$, one can compute the wavelet coefficients $s_{j} = <\psi_j,  s > $. Similarly,  one can compute a noise wavelet coefficient $n_{j} = <\psi_j, n > $ from a random noise realization $n$ or the half difference of half ring maps in Planck.\\
Choosing an isotropic wavelet function, the spherical harmonic coefficients $a^{(\psi_j)}_{l,m}$ of $ \hat{\psi_j}$ are different from zero only for $m=0$. The expected CMB power in the band $j$ is then given by :
 $$
 P_j =  \frac{1}{4 \pi}  \sum_\ell  \ell(\ell+1) \parallel  a^{(\psi_j)}_{\ell,0}  \parallel ^2   C_\ell 
 $$
where $C_\ell$ is the Planck best-fit CMB power spectrum. The local variance of the estimated CMB map in each scale at scale $j$ and pixel $k$ is performed by calculating the variance of a square patch of size $b_s \times b_s$ pixels centered on pixel $k$ in $s_j$. This procedure yields an estimate of the CMB variance map $S_j$ at each wavelet scale $j$. The same mechanism is utilized to obtain the noise variance maps $N_j$. In case of a contaminant-free CMB map, the values in the difference map $D_j =  S_j - N_j$ should be very close to the expected power $P_j$ of a pure CMB in the wavelet band $j$ up to statistical fluctuations. A strong departure of $D_j$ from $P_j$ would trace for the presence of foreground contamination.\\
We define the quality wavelet coefficient for each pixel $k$ as the ratio  $q_{j,k}  = P_j / D_{j,k}$. When $q_{j,k}$  is close to $1$, there is no statistically relevant indication for foreground contamination. Conversely, the values of $q_j,k$ close to zero suggest the presence of contaminants. The final quality map is obtained by taking the minimum of $q_j,k$ through the scales: 
$$
Q_k = \min_j  q_{j,k}
$$
Fig.~\ref{fig_qualmap} shows the quality maps for four different methods, the three official Planck maps (i.e. NILC, SEVEM, SMICA) and 
the LGMCA map obtained by jointly processing WMAP nine-year and Planck PR1 data. These maps were generated using the following command line in the open source  package {\tt iSAP} software\textcolor{red} {\footnotemark[1]}:\\

{\tt > Q = cmb\_qualitymap(CMBmap, NoiseMap, nside=2048,BS=16, NbrScale=4, Cl=Cl)}. \\

We plot 1-$Q$ rather than $Q$ which translates to reading red areas as contaminated regions. The value of $P_j$ was derived from the Planck best-fit cosmological model. It is crucial to note that the fiducial model acts only as a rescaling factor. Therefore using another model would have an extremely low impact on the figures.\\
The SMICA map having a partial sky coverage (a part of the map was inpainted), we set to zero the pixels in $Q$ which turn to be inpainted. One can observe from Figure~\ref{fig_qualmap} that SEVEM and NILC exhibit clearly much more contamination in the Galactic plane than SMICA and LGMCA. Outside the { Galactic} plane, none of these maps present significant contamination.

%===============================
 
\subsection{\label{sec:Galcenter}Galactic Center}

\begin{figure*}[htb]
\centerline{
\hbox{
\includegraphics [scale=0.45]{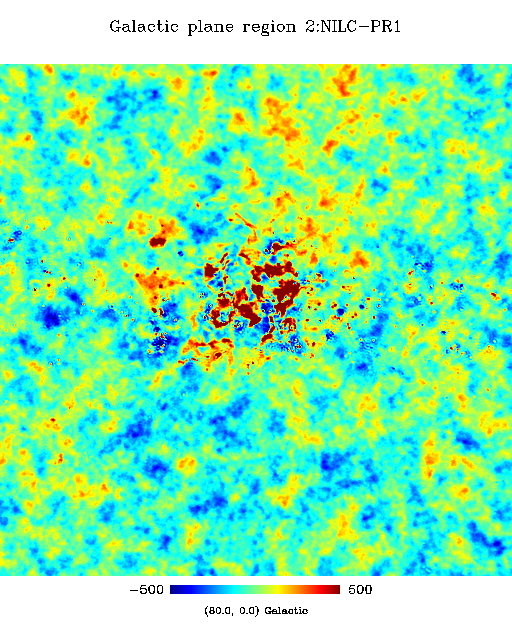}
\includegraphics [scale=0.45]{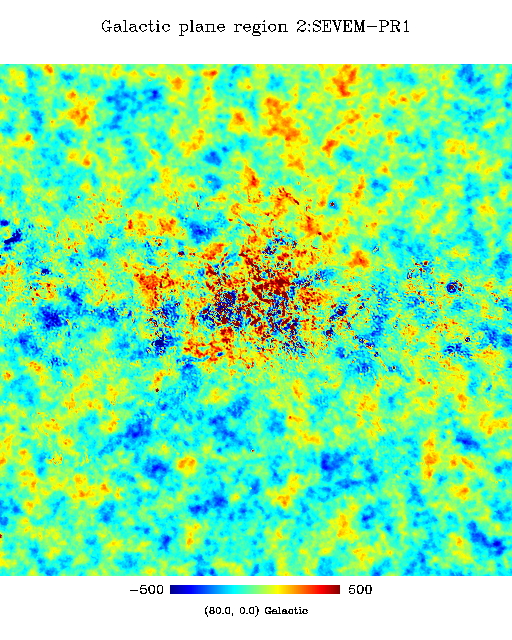}
}}
\centerline{
\hbox{
\includegraphics [scale=0.45]{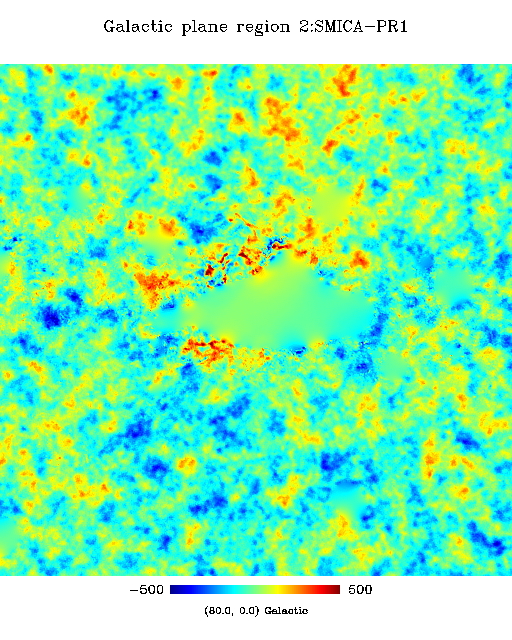}
\includegraphics [scale=0.45]{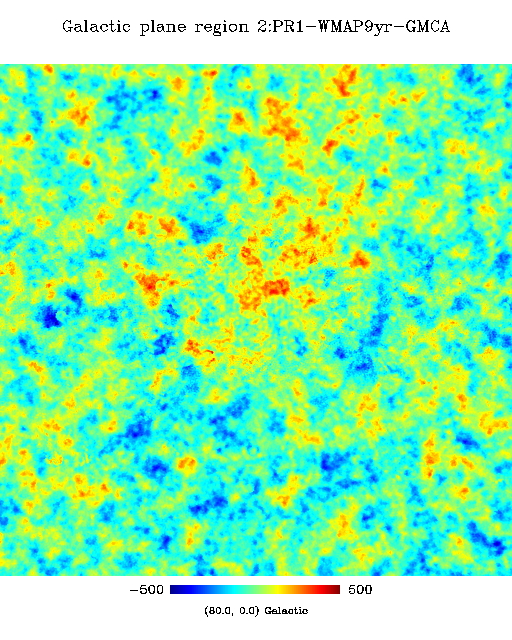}
}
}
\caption{Galactic center region, { centered at} (l,b)=(80.0,0). Top, PR1  NILC and SEVEM CMB maps, and bottom, PR1  SMICA and WPR1 LGMCA CMB maps.}
\label{fig_galcenter}
\end{figure*}

\begin{figure*}[htb]
\centerline{
\hbox{
\includegraphics [scale=0.45]{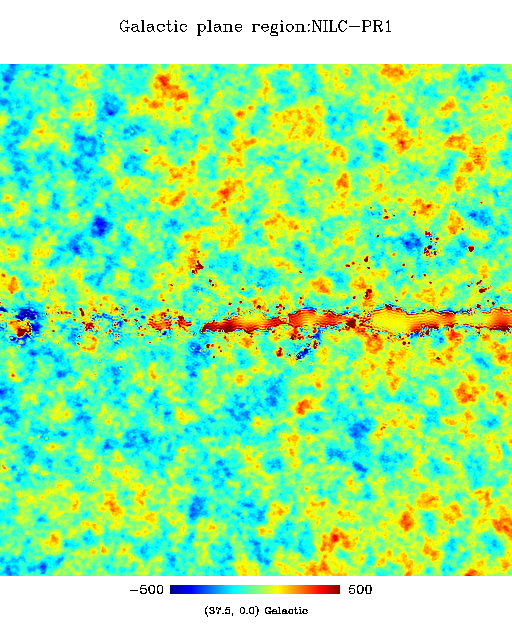}
\includegraphics [scale=0.45]{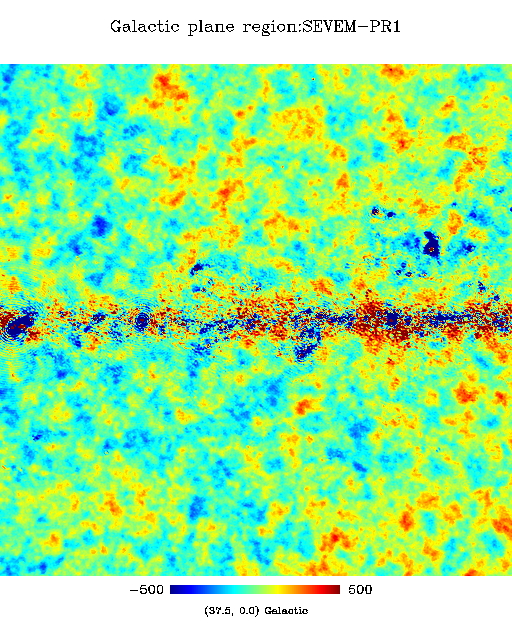}
}}
\centerline{
\hbox{
\includegraphics [scale=0.45]{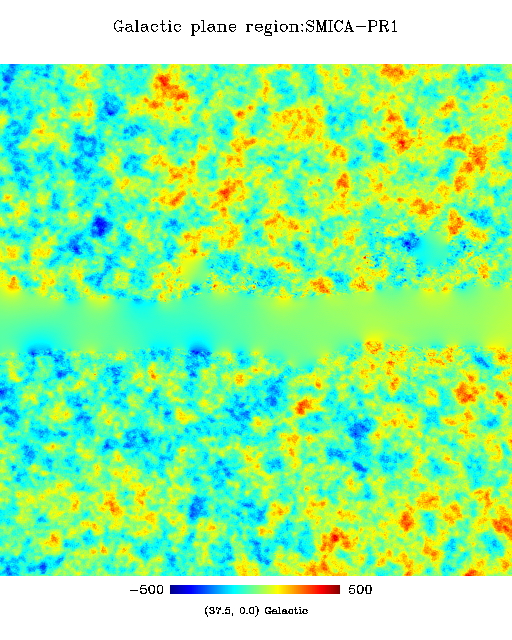}
\includegraphics [scale=0.45]{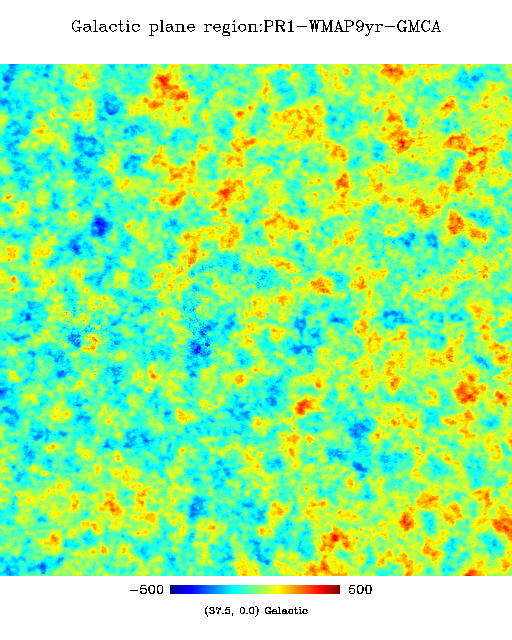}
}
}
\caption{Galactic center region, { centered at} (l,b)=(37.7,0). Top,  PR1 NILC and SEVEM CMB maps, and bottom, PR1 SMICA and WPR1 LGMCA CMB maps.}
\label{fig_galcenter2}
\end{figure*}

A glance at the CMB maps in Figure~\ref{fig:CMBmaps} is mainly significant in the { Galactic} plane. To better visualize these
differences, we show in the Fig.~\ref{fig_galcenter} and Fig.~\ref{fig_galcenter2} two regions in the { Galactic}
center. In both cases, the maps originating from the processing of Planck only all exhibit significant foreground residuals. Conversely, the LGMCA map does not present any visible remaining foreground emission. The very clean aspect of the Galactic center can be explained by the flexibility that the joint processing of WMAP and Planck affords to separate foreground components as well as the efficiency in the post-processing of the compact sources.

\subsection{\label{sec:SZ}SZ Contamination}

\begin{figure*}[htb]
\centerline{
\hbox{
\includegraphics [scale=0.3]{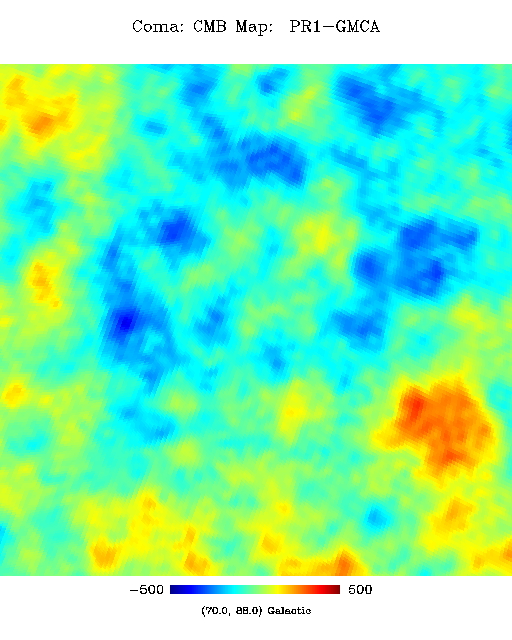}
\includegraphics [scale=0.3]{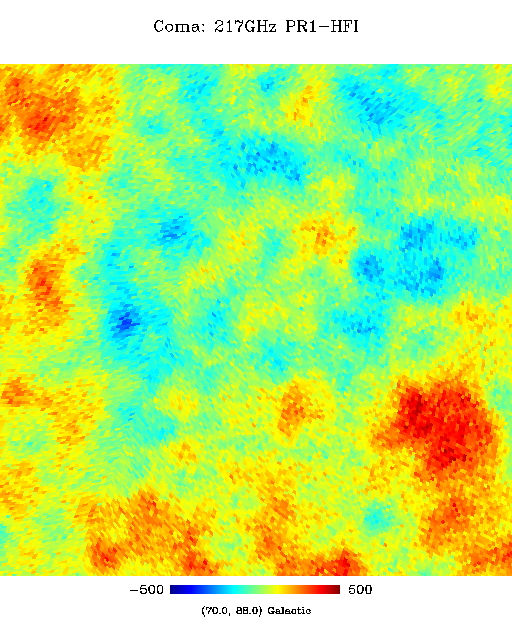}
}}
\centerline{
\hbox{
\includegraphics [scale=0.3]{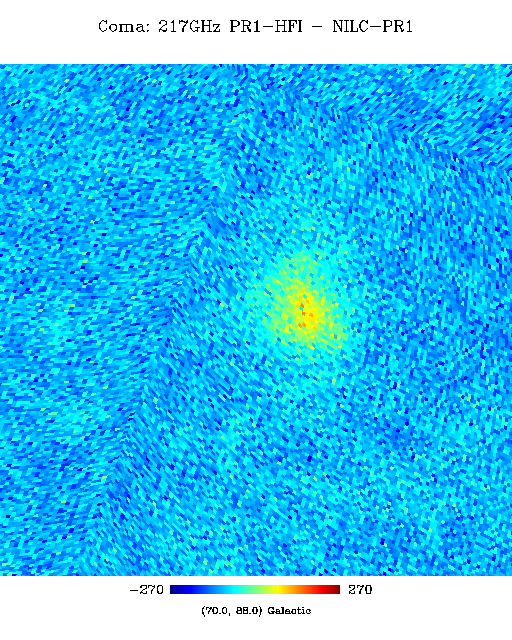}
\includegraphics [scale=0.3]{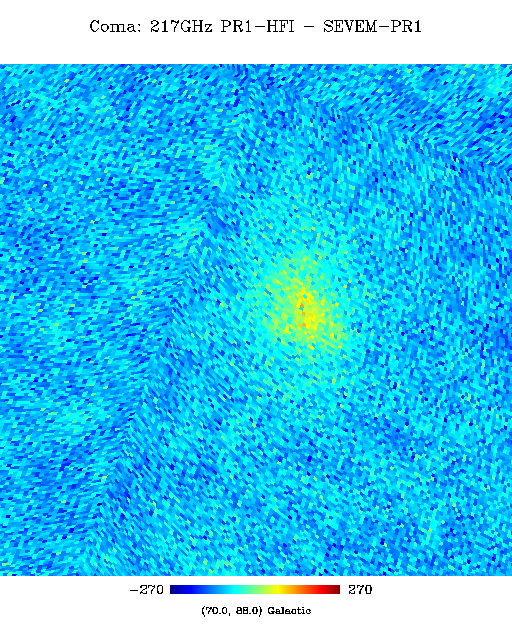}
}}
\centerline{
\hbox{
\includegraphics [scale=0.3]{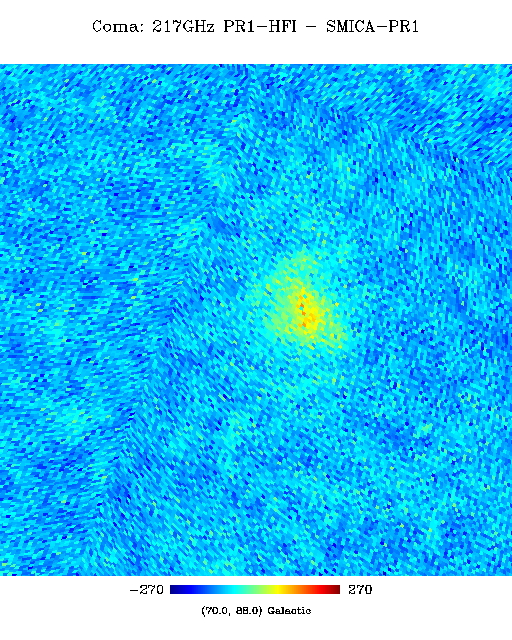}
\includegraphics [scale=0.3]{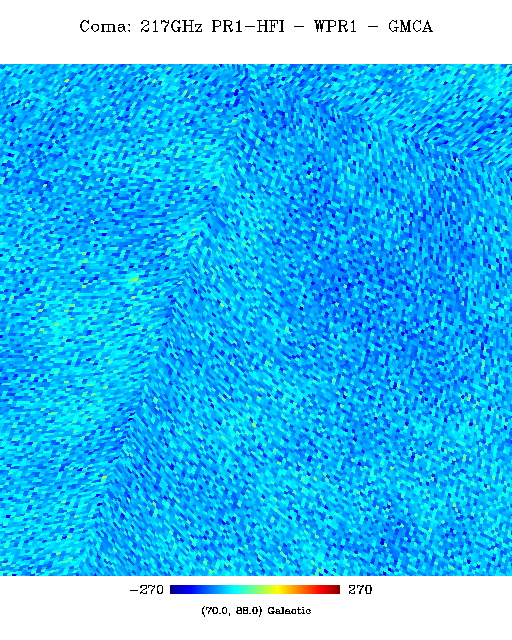}
}
}
\caption{Coma cluster area. Top, PR1 LGMCA CMB map and HFI-217 GHz map. Middle and bottom, difference map between 
HFI-217GHz and CMB maps, respectively PR1 NILC, SEVEM, SMICA and WPR1 LGMCA.}
\label{fig_coma}
\end{figure*}

\begin{figure}[htb]
\center{
\includegraphics [scale=0.25]{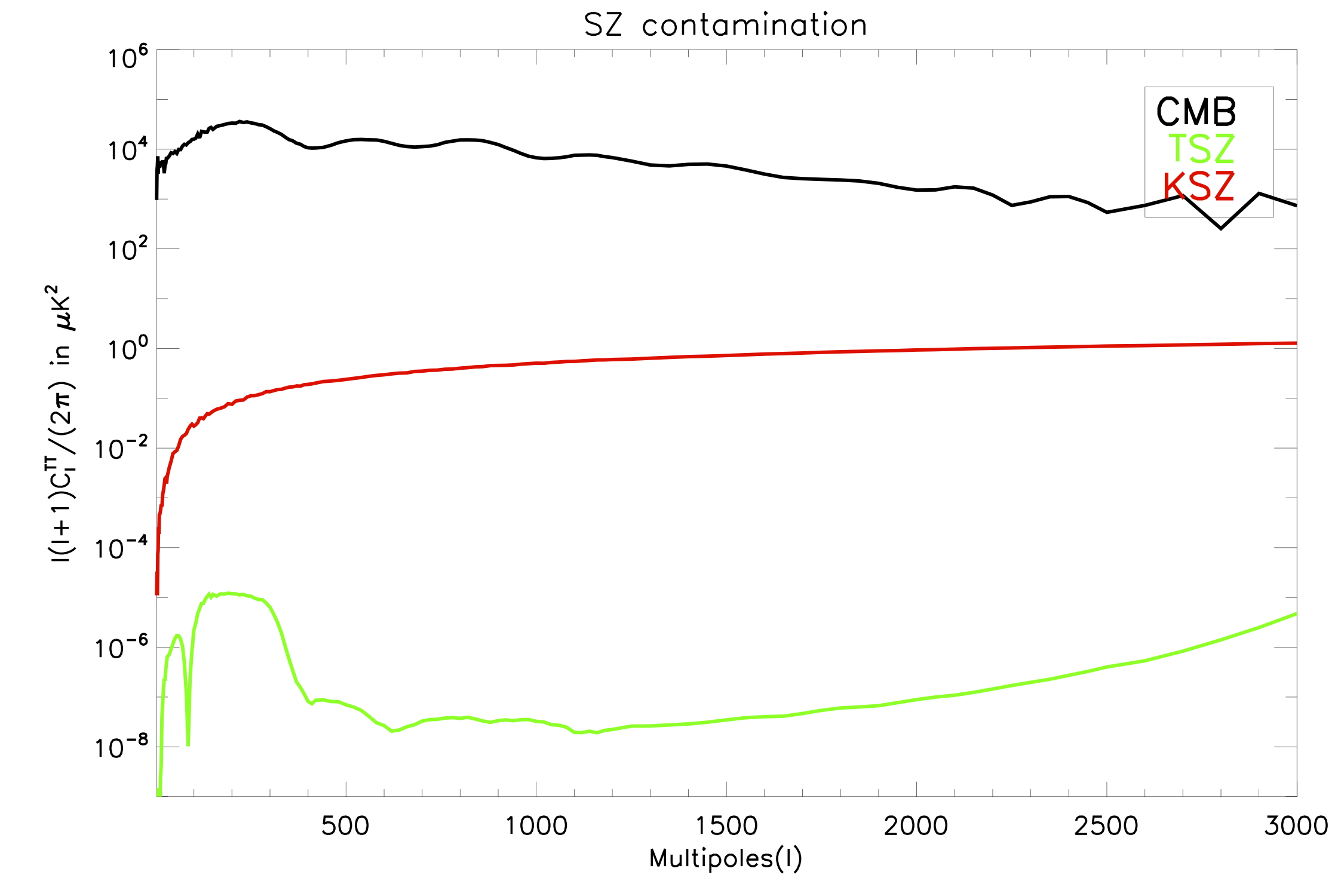}
}
\caption{tSZ and kSZ residuals in spherical harmonics in the CMB map estimated by LGMCA from simulated Planck+WMAP data.}
\label{fig_SZ_simus}
\end{figure}

The quality maps displayed in Figure~\ref{fig_qualmap} show that all the CMB maps do not exhibit significant foreground contamination outside of the Galactic center. Differences between the maps can only be measured at the { Galactic} center. At higher { Galactic} latitude, potential contamination seems to be well below the CMB fluctuations which makes them challenging to detect.\\
Fortunately, one potential foreground contamination which can be evaluated is the thermal Sunyaev Zel'Dovich (tSZ) effect. An interesting property of the tSZ effect is that it almost vanishes at $217$GHz and its contribution can considered negligible in this band. Therefore the difference between the HFI-$217$GHz channel map 
and a clean CMB map should cancel out the CMB without revealing tSZ contamination. Conversely, the same difference with a tSZ contaminated map should exhibit remaining tSZ contamination.\\
To illustrate this fact, Figure \ref{fig_coma} shows the difference maps of the four CMB maps with the HFI-$217$GHz channel map. It appears clearly that three of the maps have tSZ contamination: the Coma 
cluster can be well detected. Conversely, the LGMCA map does not present any tSZ contamination. This is expected as the 
LGMCA method takes explicitly into account the SZ emission during the component separation which therefore prevents the CMB map estimate from {being subject to} tSZ contamination. \jbc{This qualitative study can be further complemented by evaluating the level of tSZ and kSZ residuals in the CMB estimated from Planck Sky Model simulations assuming that the electromagnetic spectrum of tSZ is perfectly known. These simulations are described in Section~\ref{sec:appendix}. The contribution of the kSZ and tSZ can be estimated by applying the LGMCA parameters to the individual contribution of the kSZ and tSZ emission in the simulated frequency channels. Figure~\ref{fig_SZ_simus} displays the power spectra of the kSZ and tSZ residual contamination in the estimated CMB as well as the CMB power spectrum. Interestingly, this figure shows that tSZ residual has a contribution that can be neglected with respect to kSZ. According to these simulations, this makes the LGMCA CMB map a good candidate for kinetic SZ studies.}

% \newpage
% \clearpage

\subsection{\label{sec:Foregrounds}Assessment of the foreground contamination level}
% \begin{figure}[htb]
% \center{
%\includegraphics [scale=0.45]{../Figures/CrossCorrelPS/filters_5amin_analysis.jpg}
%}
% \caption{Filters used for residual foreground analyses.}
% \label{fig_cross_filters}
% \end{figure}
In the absence of accessible public sets of realistic simulations of the Planck sky processed by all component separation methods, assessing how foregrounds propagate to the final CMB estimates is challenging and such analysis should be performed with the highest care. The main difficulties in quantitatively comparing maps rely on: 1) a slightly different resolution for each map, 2) the masking performed on NILC and SMICA maps close to the { Galactic} center that prevents full-sky comparisons 3) the respective levels of CMB and noise make difficult to estimate residual contamination (however indicating a successful source separation over a large fraction of the sky).

To address the masking issue, all maps were first inpainted in the combined SMICA and NILC small masks (retaining $97\%$ of the sky) using the sparse inpainting technique described in \citep{inpainting:abrial06}. These maps can then be analyzed at various wavelet scales with little impact from the masked region.
The maps were degraded afterward to a common resolution of $5$ arcminutes with an additionnal low pass filter to limit the ringing of strong compact emissions. % The beams used are displayed in figure \ref{fig_cross_filters}.

\subsubsection{Cross-correlation with external templates}

\begin{figure*}[htb]
\centerline{
\hbox{
\includegraphics[scale=0.25]{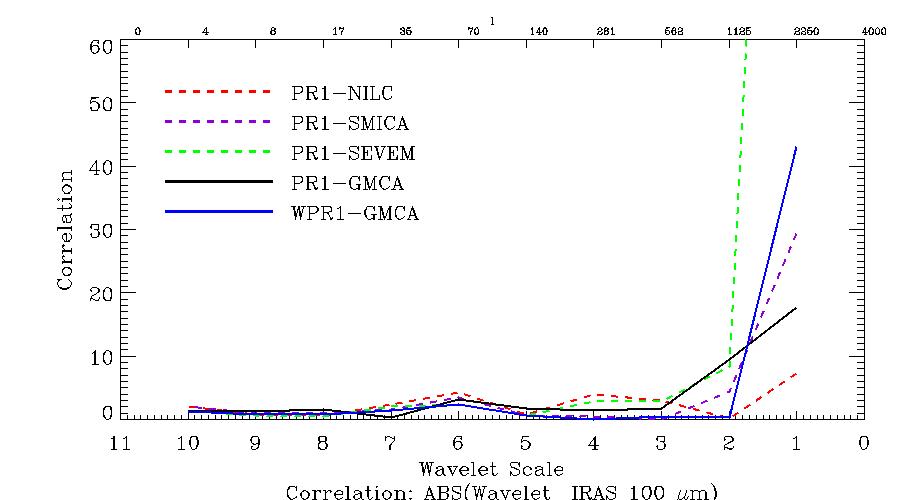}
\includegraphics[scale=0.25]{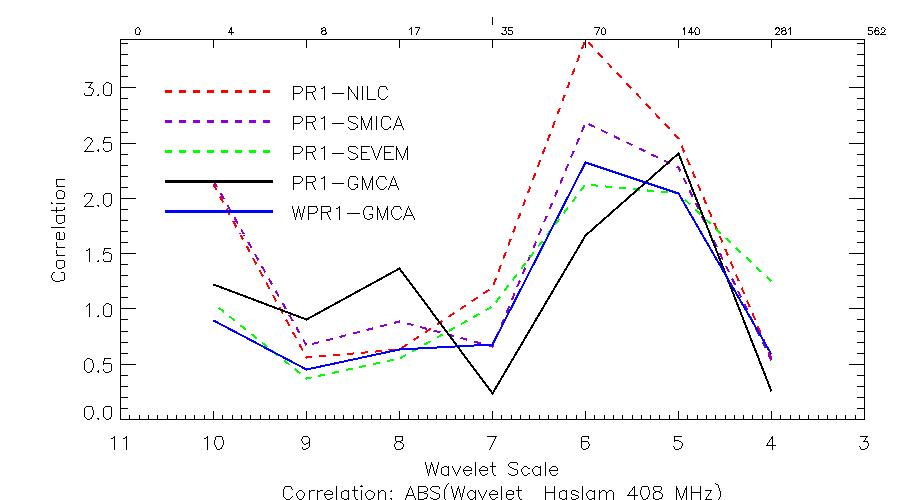}
}}
\centerline{
\hbox{
\includegraphics[scale=0.25]{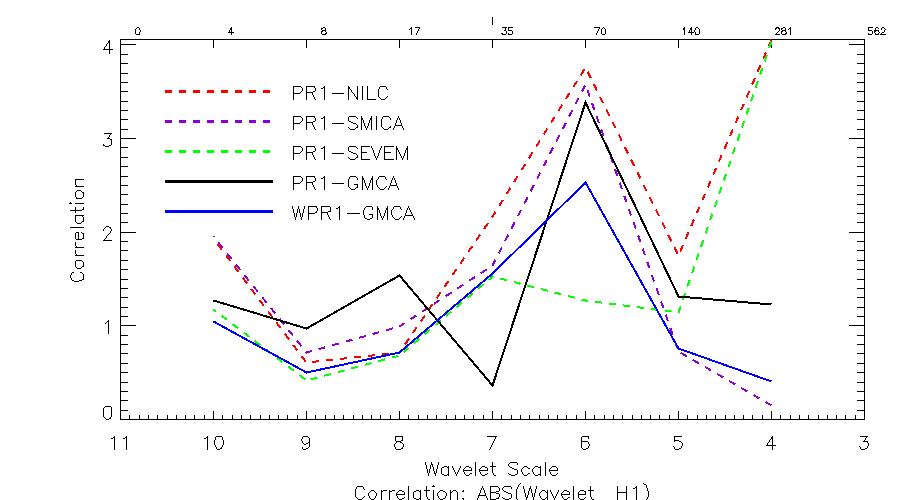}
\includegraphics[scale=0.25]{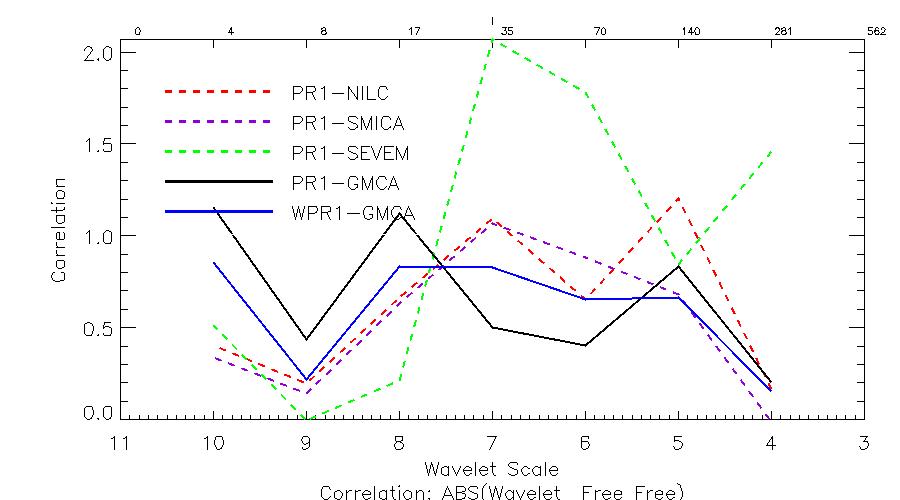}
}}
\caption{Normalized cross-correlation per wavelet scale of the CMB maps with the IRIS map (top left), the Haslam map (top right), the H1 map (bottom left) and the Free-Free template (bottom right).}
\label{fig:CorrelperWT}
\end{figure*}

Computing quantitative measures of foreground contamination from real data is not trivial. One way of measuring differences of foreground levels between CMB maps can be carried out through their cross-correlation with external templates that trace specific foreground emissions. Such cross-correlations have been performed with the Haslam $408$ MHz map, an H-alpha template provided by the WMAP consortium \citep[see][]{WMAP9_1}, a HI column density map, a velocity-integrated CO brightness temperature map (all accessible via the NASA website\footnote{{\it http://lambda.gsfc.nasa.gov/product/foreground/}}), and the IRIS $100 \mu m$ map \citep{IRIS05}. A set of 80 realizations of CMB maps (assuming the fiducial cosmological model obtained from the Planck PR1 results) and noise maps were also processed through the LGMCA pipeline. 
Statistics obtained from the CMB maps were normalized by the standard deviation of the 
statistics computed from the noise realizations. As these noise realizations are only valid for the WPR1 LGMCA map (in the absence of similar propagated noise realizations for other component separation methods available), this normalized correlation should be understood as a rescaling between the 
different wavelet scales, and the amplitude only as a very rough approximation of the correlation SNR.  
Fig.\ref{fig:CorrelperWT} shows the normalized correlation for different CMB maps, the three
released Planck PR1 maps (i.e. NILC, SMICA, SEVEM), the WPR1 LGMCA map
and the PR1 LGMCA map. The latest is shown in order to see whether adding WMAP-9yr channels improves the cross-correlation with the different templates.
We can see that none of these plots exhibit statistically significant cross-correlations ({\it i.e.} normalized correlation larger than five), with the exception of the finest scale in 
the one calculated with the IRIS map. This cross-correlation could be attributed to dust or CIB.
Investigating more this effect, we have seen that this strong correlation at fine scales in the SEVEM map disappears when we use the 70\% fsky mask provided with the SEVEM map, which means it could be due to either unmasked infrared-point sources in the small combined SMICA and NILC mask, or 
residual dust in the Galactic center which remain in the SEVEM map. Concerning the WPR1 LGMCA map, the same experience did not remove this effect, so the contamination is most likely due to CIB.

\subsubsection{Higher order statistics}
\label{sec:hos}
\begin{figure*}[htb]
\centerline{
\hbox{
\includegraphics [scale=0.25]{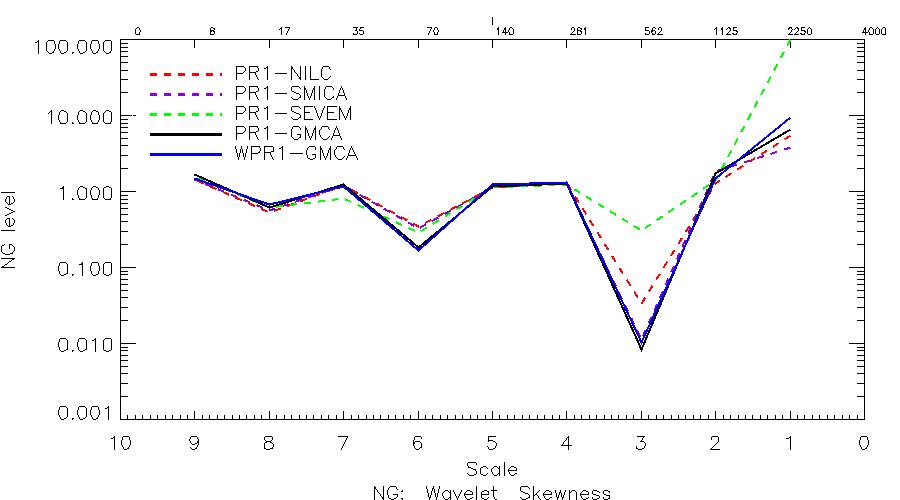}
\hspace{0.5cm}
\includegraphics [scale=0.25]{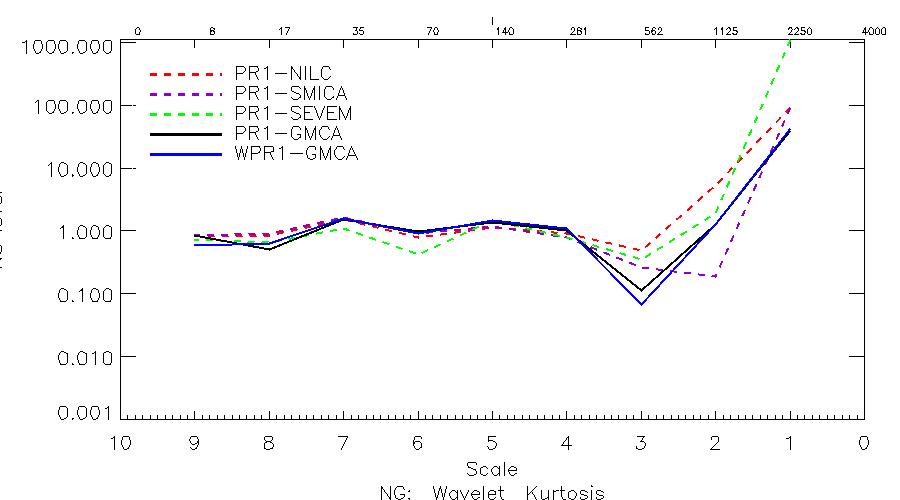}
}
}
\centerline{
\hbox{
\includegraphics [scale=0.25]{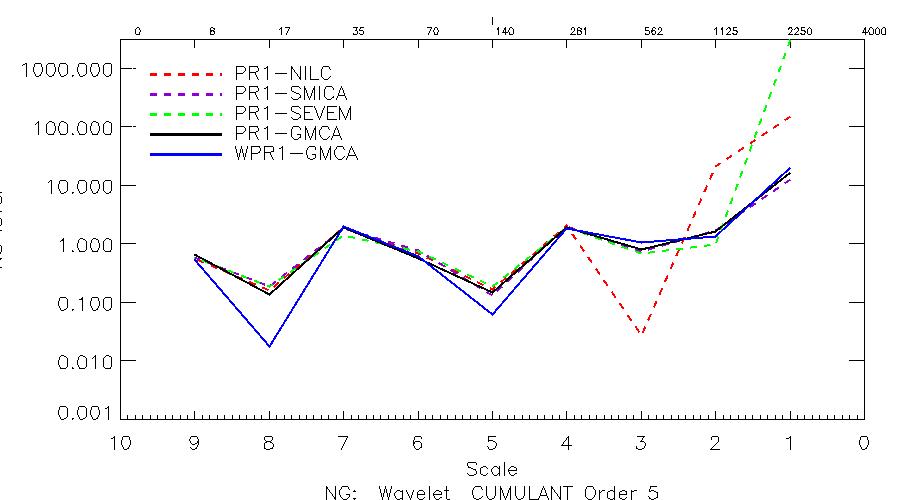}
\hspace{0.5cm}
\includegraphics [scale=0.25]{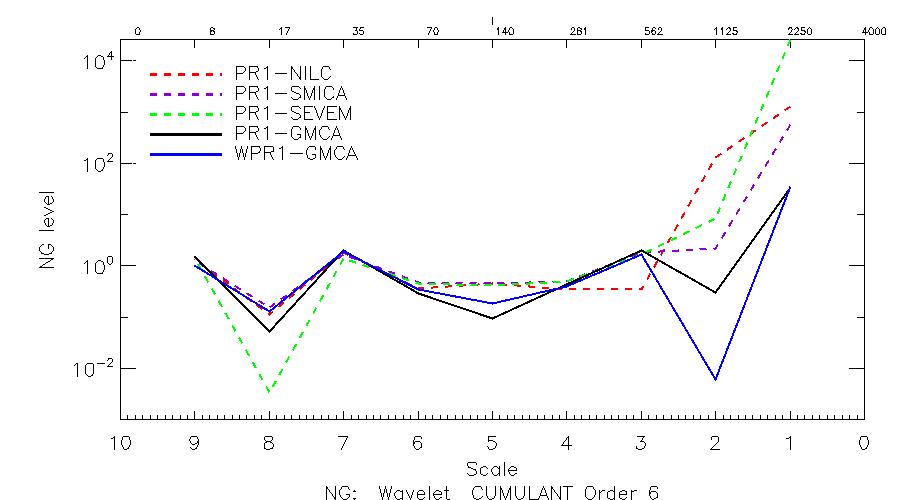}
}
}
\caption{Normalized high order statistics computed at various wavelet scales for the high resolution masks inside the inpainting mask (about $97 \%$ of the sky): Skewness (top left), Kurtosis (top right), cumulant of order 5 and 6 (bottom left and right). }
\label{fig_HOS}
\end{figure*}

\begin{figure*}[htb]
\centerline{
\hbox{
\includegraphics [scale=0.25]{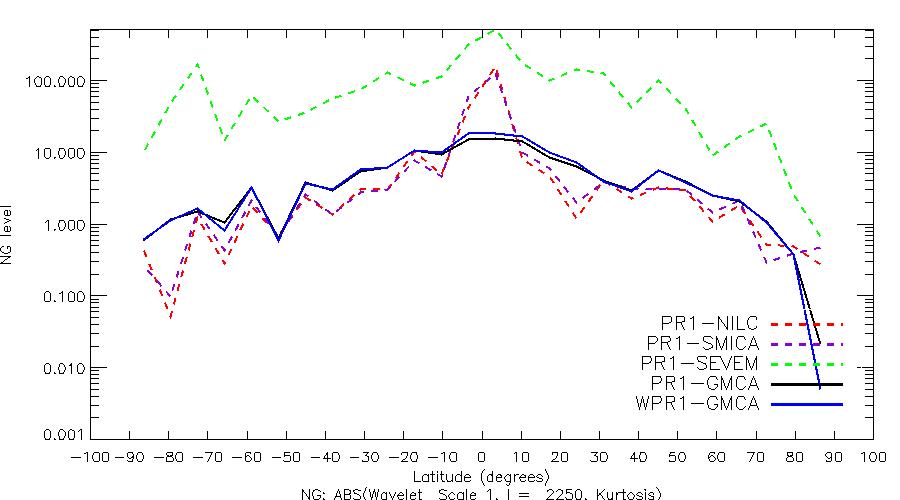}
\hspace{0.5cm}
\includegraphics [scale=0.25]{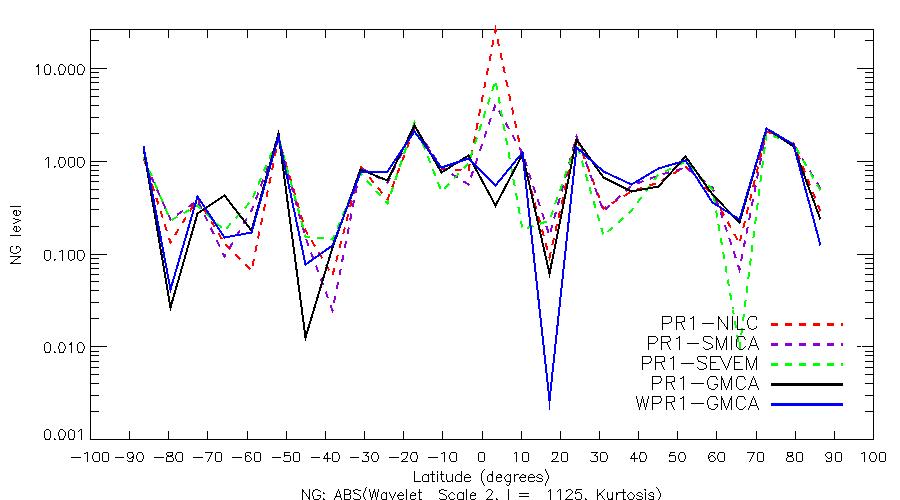}
}
}
\centerline{
\hbox{
\includegraphics [scale=0.25]{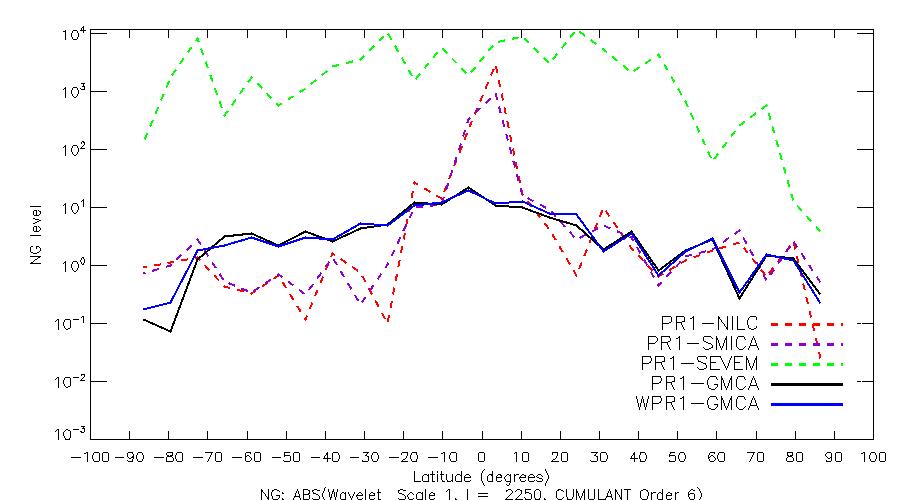}
\hspace{0.5cm}
\includegraphics [scale=0.25]{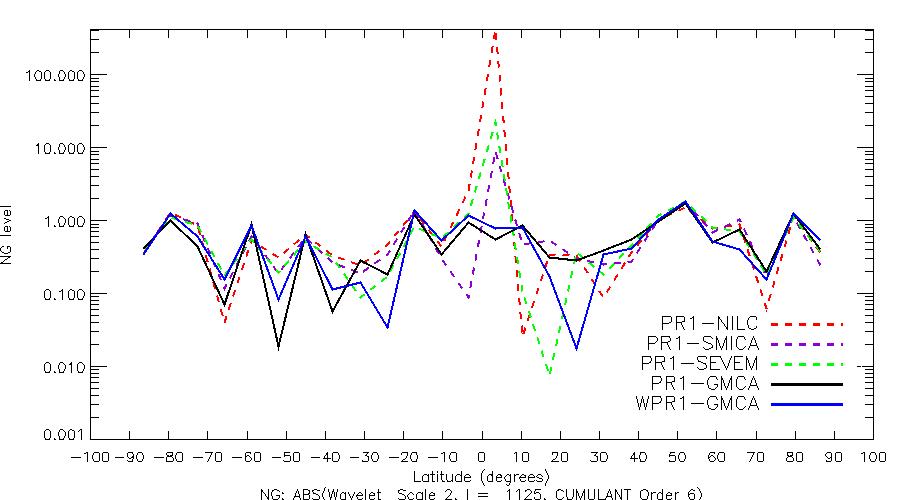}
}
}
\caption{Normalized high order statistics pet latitude band computed from the two finest wavelet scales for the high resolution masks inside the inpainting mask (about $97 \%$ of the sky): top, normalized Kurtosis at scale 1 and 2, and 
bottom, normalized cumulant of order 6 at scale 1 and 2. }
\label{fig_scales_1and2_HOS}
\end{figure*}

Deviations from Gaussianity is another way to quantify the level of remaining foregrounds without requiring ancillary templates. For that purpose, higher-order statistics provide a model-independent measure of non-Gaussianity (NG) which can be further enhanced when evaluated in the wavelet domain. Figure~\ref{fig_HOS} shows the skewness,
the kurtosis, and the cumulants of order 5 and 6. These values are computed using the same sky coverage as in the previous section, and are normalized in a similar way. We can see that strong departures from Gaussianity is observed
in the first two scales. To better characterize these NG, we have also performed the same analysis but per latitude band.
Figure~\ref{fig_scales_1and2_HOS} top shows the normalized kurtosis at scales 1 and 2, and Figure~\ref{fig_scales_1and2_HOS} bottom shows the normalized cumulant of order 6 at scales 1 and 2.\\

From this evaluation, we can conclude that:
\begin{itemize}
\item All maps are compatible with the Gaussianity assumption up to the second wavelet scale. 
\item LGMCA maps (PR1 and WPR1) present the best behavior at the finest scales. This is also an indication that
the contamination shown in the previous detection is rather due to CIB, since dust contamination would have certainly 
impacted the high order statistics.
\item  Deviation from Gaussianity is significant for all maps, but only in the finest scales. These NG are
clearly due to foreground residuals in the { Galactic} plane, except for SEVEM where point sources contaminate the map more strongly and need to be masked. This confirms that a much more conservative mask is required for cosmological non-Gaussianity studies, especially when the finest scales are used such as for CMB weak lensing or $f_{nl}$ detection.
\item Point source processing methods that were used for all maps, except SEVEM, seem to work properly outside the 
{ Galactic} plane, since no NG are detected.
\end{itemize}

\subsubsection{WPR1 versus PR1}
The differences between the WPR1 LGMCA map and the currently available maps can have many origins which go beyond the joint processing with the WMAP data : i) sparsity is used as a separation criterion, ii) post-processing of the point sources and the Galactic center. From our tests, the differences between the WPR1 LGMCA and PR1 LGMCA maps, obtained by performing LGMCA on the Planck data only, show relatively 
few significant differences. The main one concerns the { correlation with the H1} template, where at scale $6$, the PR1 map presents a significant cross-corrrelation with the H1 template (about $4 \,\sigma$ detection) while this quantity remains insignificant for the WPR1 LGMCA map.\\
Another interesting aspect relative to the joint reconstruction is the fact that residual systematics in Planck and WMAP are likely to be unrelated. From this perspective, we can expect that combining the Planck and WMAP data would lead to a CMB with less systematics \citep{2013arXiv1303.5083P,Frejsel,Naselsky2012, Gruppuso2013}. This should be especially the case for the largest scales of the CMB map where correcting for the systematic effects is particularly challenging. This suggests that a study of the large scale multipoles of the CMB map should be thoroughly performed which we leave for future work.

%%%%%%%%%%%%%%%%%%%%%%%%%%%%%%%%%%%%%%%%%%%%%%%%%%%%%%%
\section{\label{sec:repres}Reproducible Research}

\begin{table*}[htbp]
  \centering
  \begin{tabular}{@{} lcl @{}} % Column formatting, @{} suppresses leading/trailing space
  Product Name  &Type & Description\\
  \hline
  Planck WPR1 products:\\
  WPR1\_CMB\_muK\_hr1.fits   &  Map  & WPR1 CMB estimate, first half ring  \\
  WPR1\_CMB\_muK\_hr2.fits   &  Map  & WPR1 CMB estimate, second half ring\\
  WPR1\_CMB\_muK.fits   &  Map  & WPR1 CMB map estimate \\
  WPR1\_CMB\_noise\_muK.fits   &  Map  & WPR1 noise map estimate \\
  WPR1\_CMB\_rms\_muK.fits   &  Map  & WPR1 root mean square error of the CMB map estimate (see Section \ref{sec:rmsmap}) \\
  PR1\_CMB\_muK\_hr1.fits   &  Map  & PR1 CMB estimate, first half ring  \\
  PR1\_CMB\_muK\_hr2.fits   &  Map  & PR1 CMB estimate, second half ring\\
  PR1\_CMB\_muK.fits   &  Map  & PR1 CMB map estimate \\
  PR1\_CMB\_noise\_muK.fits   &  Map  & PR1 noise map estimate \\
  \hline 
  Software products:\\
  {\tt run\_lgmca\_wpr1\_getmaps.pro}  & code (IDL)  & code to compute the CMB map estimates\\ & & (requires {\tt HealPix} and {\tt iSAP}). \\
  {\tt wpr1\_analysis\_routines.pro}  & code (IDL)  & routines to reproduce the figures of the paper. \\ & & (requires {\tt HealPix} and {\tt iSAP}).\\
%  {\tt wmap9\_remove\_point\_sources.pro}  & code  (IDL)  & script which applies SPSR to all WMAP9 channels\\ & & (requires {\tt HealPix} and {\tt iSAP}).\\
  \hline

  \end{tabular}
  \caption{List of products made available in this paper in the spirit of reproducible research, available here: {http://www.cosmostat.org/planck\_wpr1.html}.}
  \label{tab:reproducible_ps}
\end{table*}
% \citep{RepRes09}
In the spirit of participating in reproducible research, we make all codes and resulting products which constitute main results of this paper public. In Table \ref{tab:reproducible_ps} we list all such products which are made freely available as a result of this paper, and which are available here: {http://www.cosmostat.org/planck\_wpr1.html}.

%\section*{Code and products}
%The CQA code will be available at \url{http://www.cosmostat.org/lgmca}.
%LGMCA CMB maps (Planck only and  joint WMAP-Planck reconstruction) and the estimated power spectrum will also be available at \url{http://www.cosmostat.org/product}.

%%%%%%%%%%%%%%%%%%%%%%%%%%%%%%%%%%%%%%%%%%%%%%%%%%%%%%%%%%%%%%%%%%
\section{Conclusion}
We combined the WMAP nine-year and Planck PR1 data to produce a clean full-sky CMB map (without inpainted or interpolated pixels). The joint processing  of the WMAP and Planck is carried out by a recently introduced sparsity-based component separation method coined LGMCA. It also benefits from an effective post-processing of point sources based on the MCA. We show that this processing yields a full-sky CMB map with no significant foreground residuals on the Galactic center. Moreover, the larger number of d.o.f.'s { due to} the joint processing of WMAP and Planck  allows for the estimation of a map without detectable tSZ contamination, in contrast to existing available CMB maps. 
%The LGMCA CMB map has a very low correlation with the IRIS map for large and intermediate scales. It exhibit slightly higher correlations at the finest scales which are likely due to a higher contamination from CIB. Similarly to the currently available maps, the LGMCA map  exhibits no significant departure from gaussianity outside the galactic center. .... etc
Our conclusions relative to the WPR1 LGMCA CMB map are:
\begin{itemize}
\item it is the only one which is full-sky, without requiring any inpainting techniques.
\item it is virtually free of tSZ contamination, so it should be the best candidate for the kSZ studies.
\item it is very clean, even in the { Galactic} plane.
\item it presents however a slightly higher CIB contamination detectable for $l> 2000$.
\item assuming the power spectrum of the LGMCA CMB map are similar has error bars which are similar to those estimated by the Consortium, taking into account 
all instrumental effects and residual foregrounds (beam uncertainties, point sources, CIB, etc), the GMCA estimated power spectra does not show discrepancy 
with the Planck best-fit power spectrum.
\item the WPR1 LGMCA map present a slightly lower level of H1 contamination than the PR1 map while exhibiting no statistically significant differences otherwise.
\end{itemize}
This suggests that further study should emphasize on the analysis of the large scale structure of the CMB from the PR1 LGMCA and WPR1 LGMCA maps.

\begin{acknowledgements}
We would like to thank the anonymous reviewer for his comments that greatly helped improving the paper. 
This work is supported by the European Research Council grant SparseAstro (ERC-228261), the Centre National des Etudes Spatiales (CNES),
and by the Swiss National Science Foundation (SNSF).
 We used Healpix software \citep{Healpix},
  iSAP\footnote{{http://jstarck.free.fr/isap.html}} software, \emph{WMAP} data \footnote{{http://map.gsfc.nasa.gov}},  and Planck data  \footnote{{http://www.sciops.esa.int/wikiSI/planckpla}} .
\end{acknowledgements}

\bibliographystyle{aa} 
\bibliography{lgmca_wpr1}

\vspace{-.5cm}
\appendix

\section{Simulations}
\label{sec:appendix} 

\subsection*{Simulated data}

\jbc{The LGMCA algorithm is applied to the WMAP and Planck data, which is simulated by the Planck Sky Model (PSM)\footnote{For more details please visit the PSM website: {\it http://www.apc.univ-paris7.fr/~delabrou/PSM/psm.html}.} \citep{PSM12}. The PSM models the instrumental noise, the beams and the astrophysical foregrounds in the frequency range that is probed by WMAP and Planck. The simulations were obtained as follows;
\begin{itemize}
\item{\bf Frequency channels:} the simulated data contains the $14$ WMAP and Planck frequency channels ranging from $23$ to $857$GHz. The frequency-dependent beams are assumed to be isotropic Gaussian PSFs.
\item{\bf Instrumental noise:} instrumental noise has been generated according to a Gaussian distribution, with a covariance matrix provided by the WMAP (9-year) and the Planck consortia.
\item{\bf Cosmic microwave background:} the CMB map is drawn from a Gaussian random field with WMAP 9-year best-fit theoretical power spectrum (from the $6$ cosmological parameters model). No non-Gaussianities, such as lensing or ISW effects, have been added to the CMB map. 
\item{\it Synchrotron:} this emission arises from the acceleration of the cosmic-ray electrons in the magnetic field of our Galaxy. It follows a power law with a spectral index that varies across pixels from $-3.4$ and $-2.3$ \citep{gauss:bennett03}. In the Planck data, this component mainly appears at lower frequency observations ({\it typ.} $\nu < 70$GHz).
\item{\it Free-Free:} the free-free emission is due to the electron-ion scattering and follows a power law distribution with an almost constant spectral index across the sky ($\sim-2.15$) \citep{Dick03}. 
\item{\it Dust emission:} this component arises from the thermal radiation of the dust grains in the Milky Way. This emission follows a gray body spectrum which depends on two parameters: the dust temperature and the spectral index \citep{Fink99}. Recent studies, involving the joint analysis of IRIS and Planck $545$ and $857$Ghz observations, show significant variations in both the dust temperature and the spectral index across the sky both on large and small scales \citep{PER_Dust}.
\item{\it AME:} the AME (anomalous microwave emission) -- or spinning dust -- may develop from the emission of spinning dust grains on nanoscales. This component has a spatial correlation with the thermal dust emission but has an emissivity that roughly follows a power law in the frequency range of Planck and WMAP \citep{PER_SDust}.
\item{\it CIB:} cosmological infra-red background originates from the emission of unresolved galaxies at high redshifts.
\item{\it CO:} CO emission has been simulated using the DAME H1 line survey \citep{Dame01}.
\item{\it SZ:} the Sunyaev-Zel'Dovich effect results from the interactions of the high energy electrons and the CMB photons through inverse Compton scattering \citep{SZoriginal}. The SZ electromagnetic spectrum is well known to be constant across the sky. 
\item{\it Point sources:} these components belong to two categories of radio and infra-red point sources, which can be of Galactic or extra-Galactic origins. Most of the brightest compact sources are found in the ERCSC catalogue provided by the Planck mission \citep{PER_ERCSC}. These point sources have individual electromagnetic spectra.
\end{itemize}
The same IRIS $100 \mu m$ map and parameters, as listed in Table~\ref{tab_resol_planck}, were used for LGMCA.}

\subsection*{Recovered Maps}
\begin{figure}[htb] 
\centerline{\includegraphics [scale=0.25]{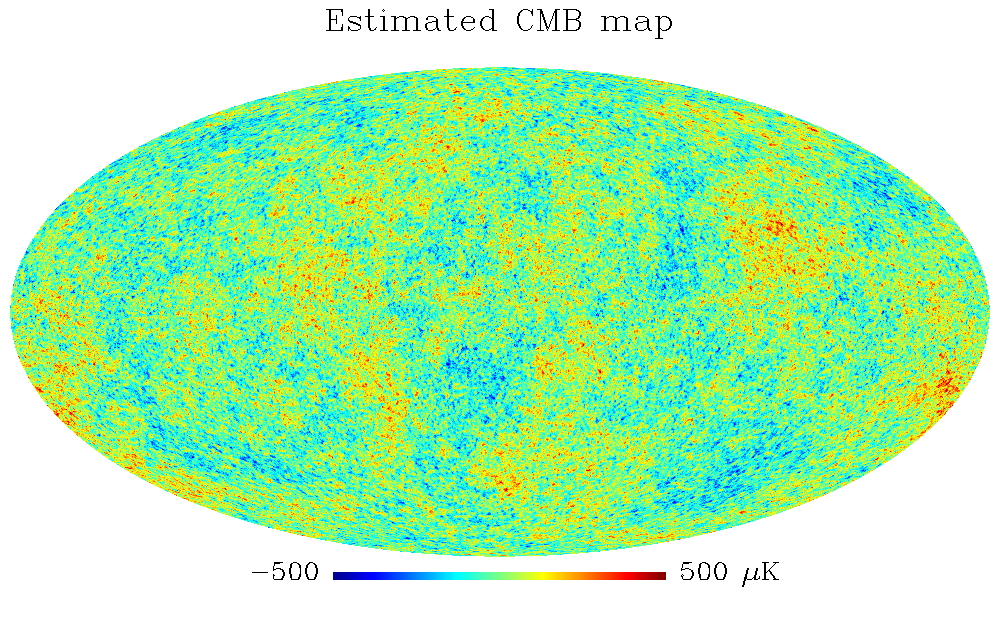}}
\caption{CMB estimated from simulated WMAP and Planck data.}
\label{fig_simus_CMB}
\end{figure}

\begin{figure}[htb]
\centerline{\includegraphics [scale=0.25]{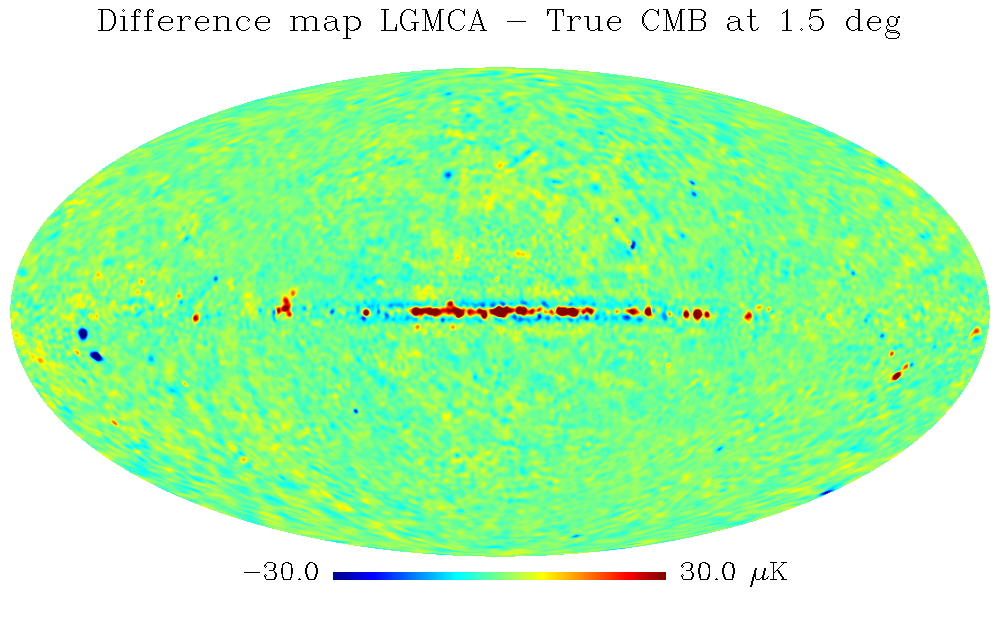}}
\caption{Residual map defined as the difference between the estimated CMB and the true simulated CMB map at resolution $1.5$ degree.}
\label{fig_simus_CMB_error}
\end{figure}

\begin{figure*}[htb]
\centerline{
\hbox{
\includegraphics[scale=0.2]{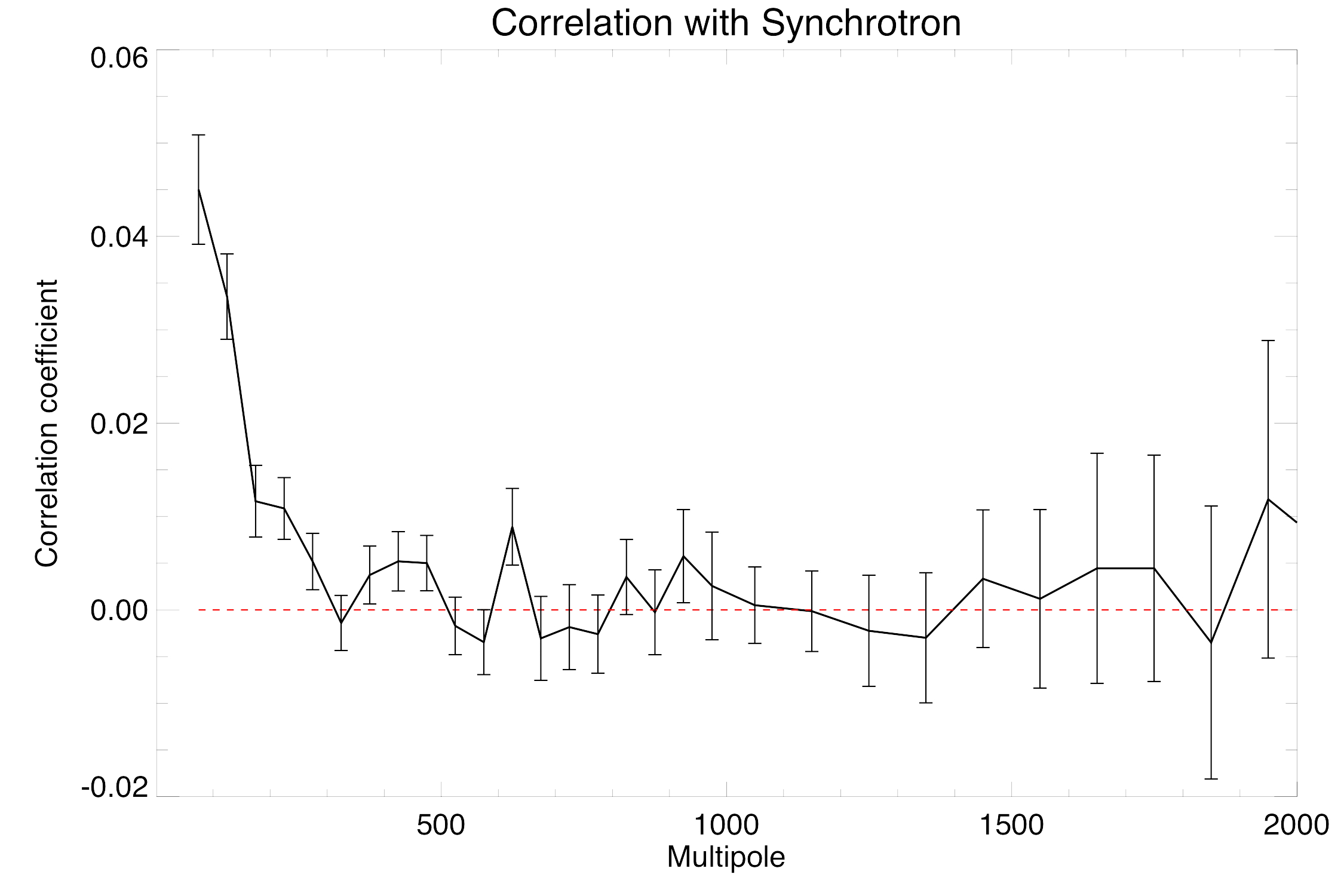} \hspace{0.2cm} 
\includegraphics[scale=0.2]{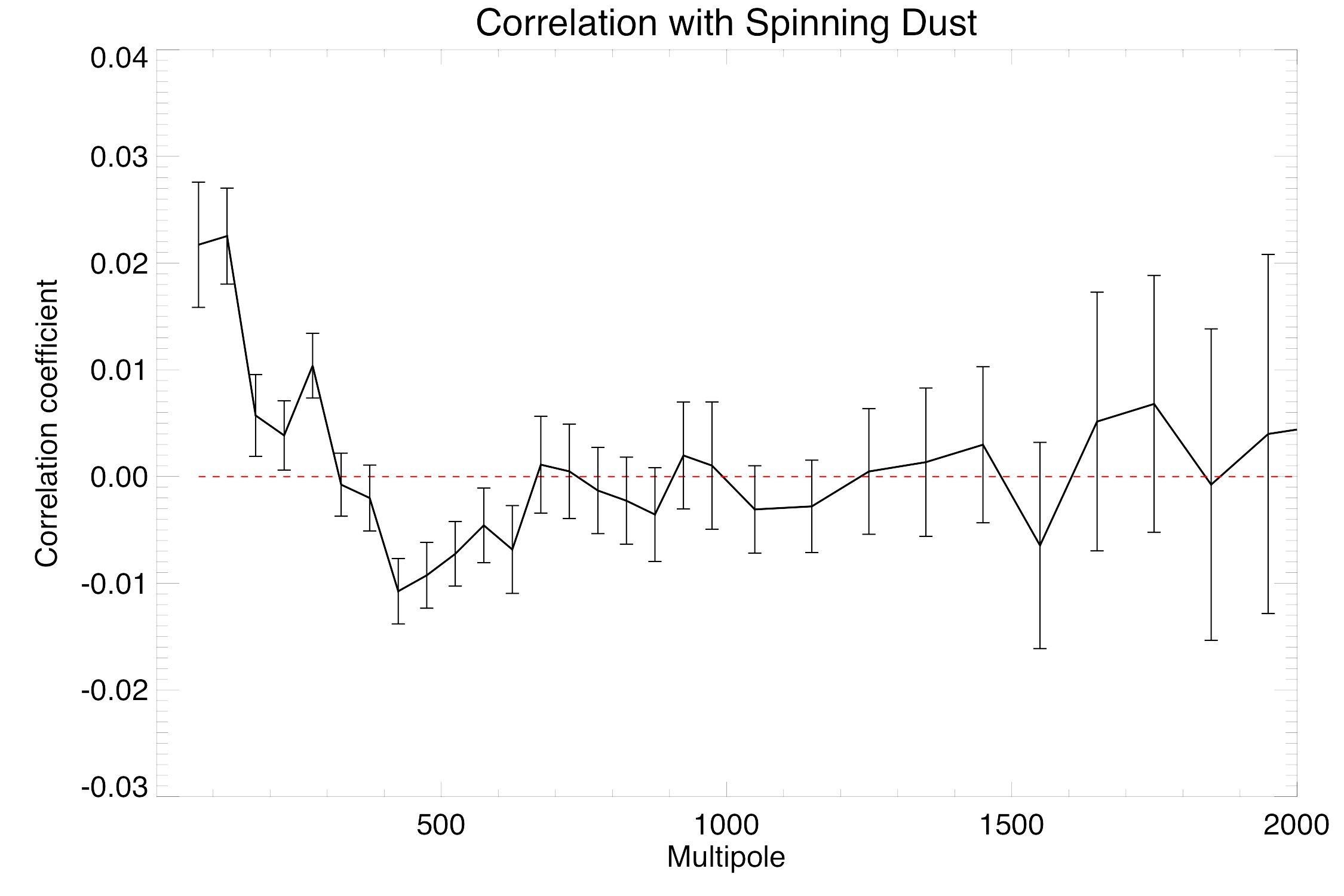}
}}
\centerline{
\hbox{
\includegraphics[scale=0.2]{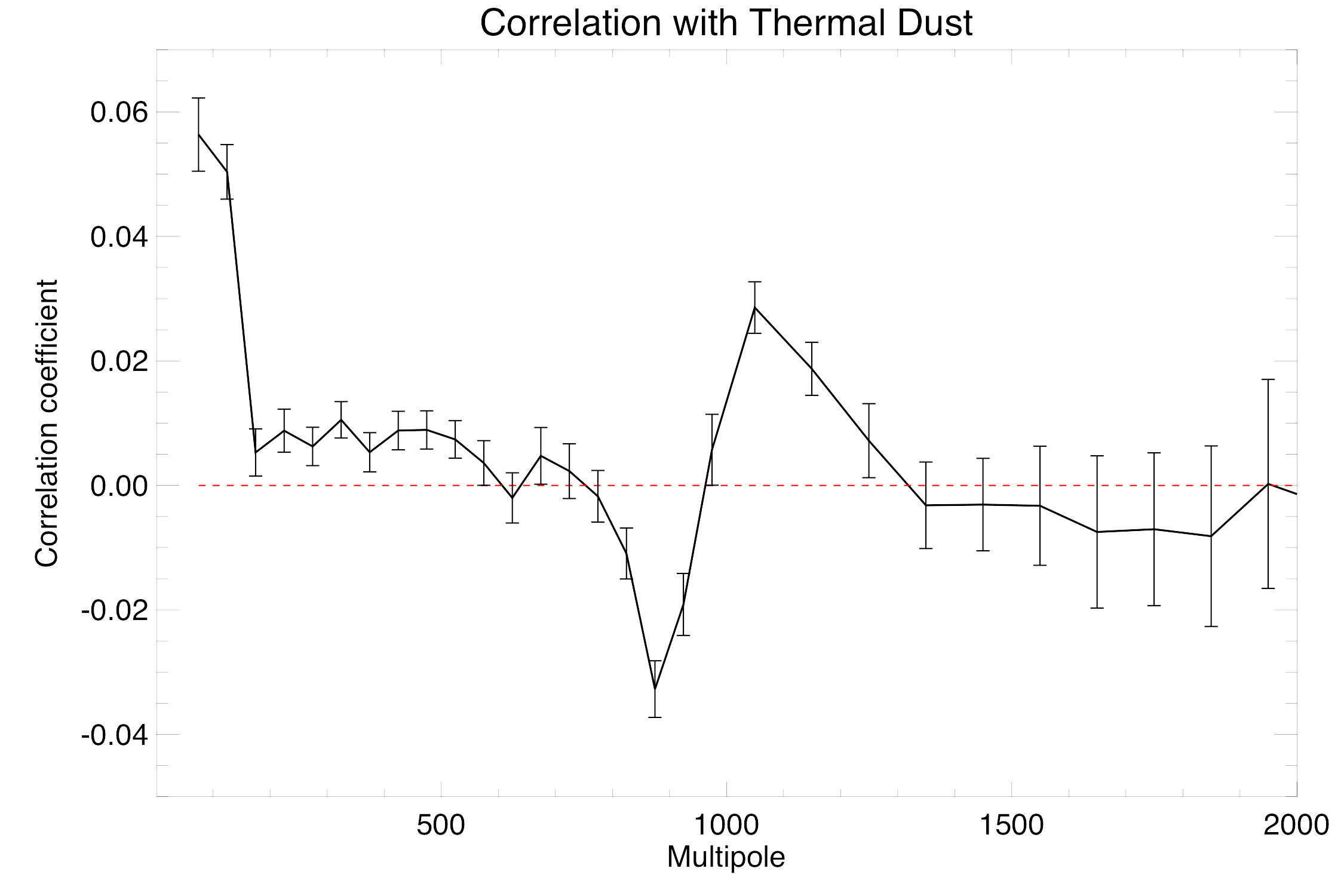} \hspace{0.2cm} 
\includegraphics[scale=0.2]{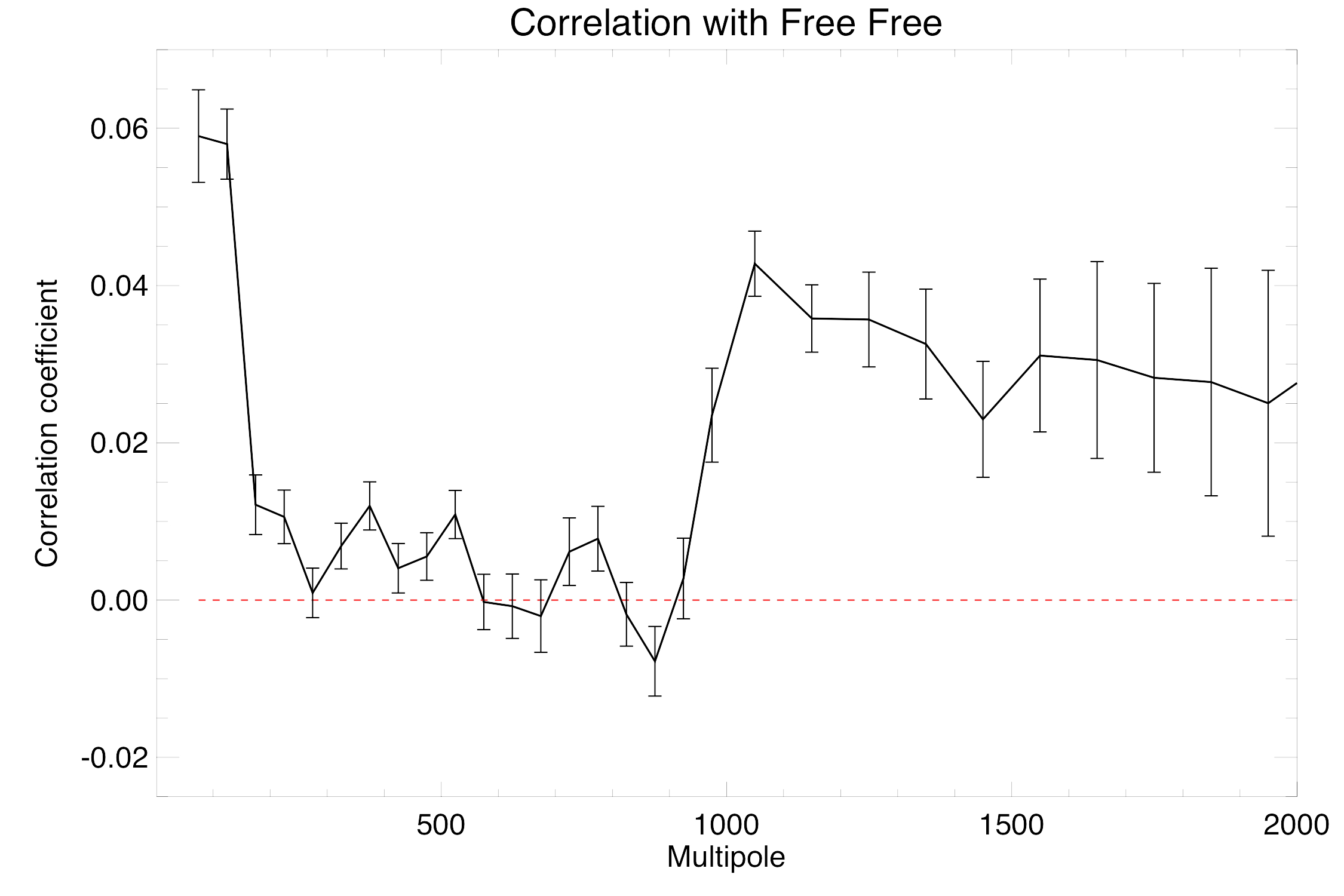}
}}
\centerline{
\hbox{
\includegraphics[scale=0.2]{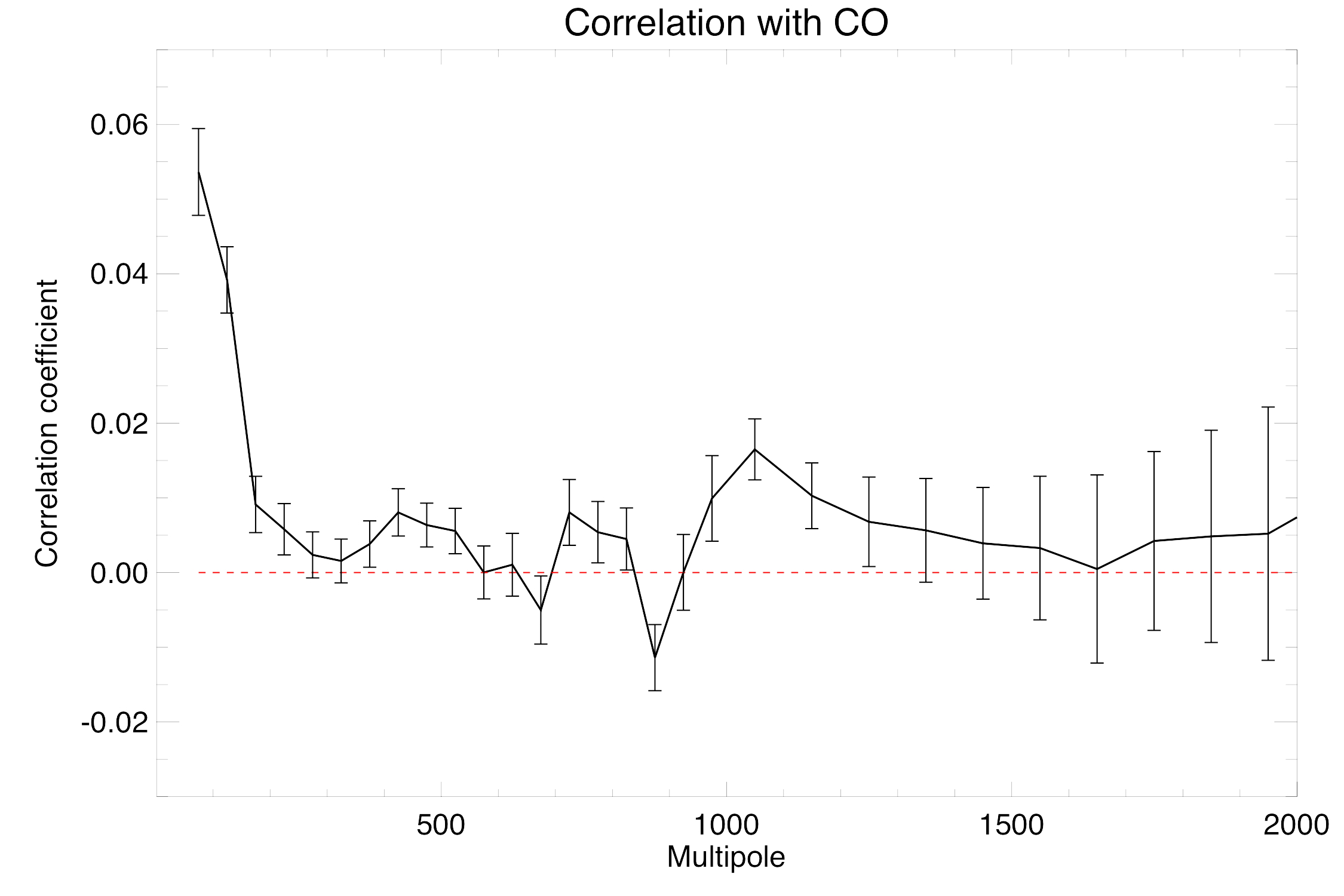} \hspace{0.2cm} 
\includegraphics[scale=0.2]{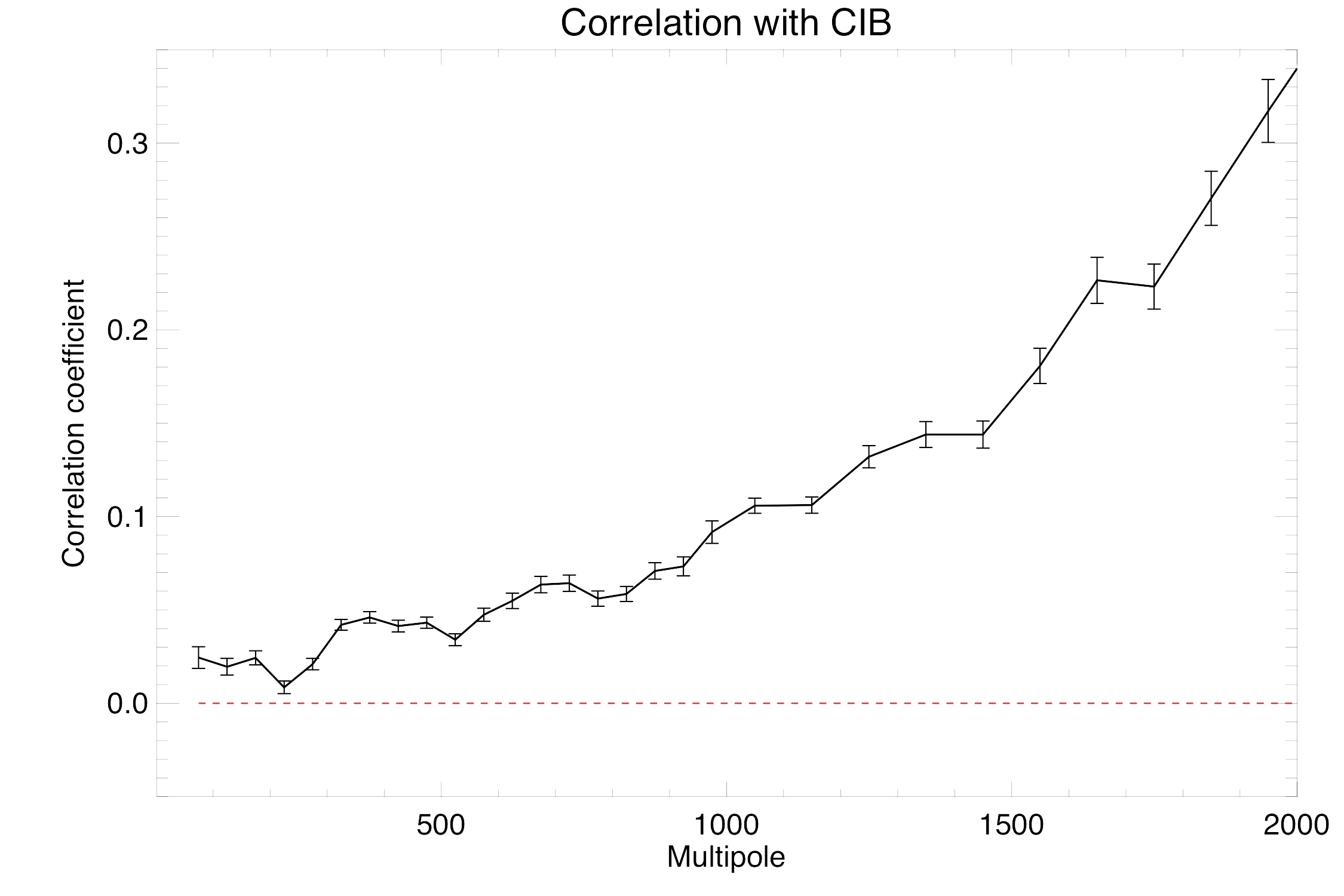}
}}
\centerline{
\hbox{
\includegraphics[scale=0.2]{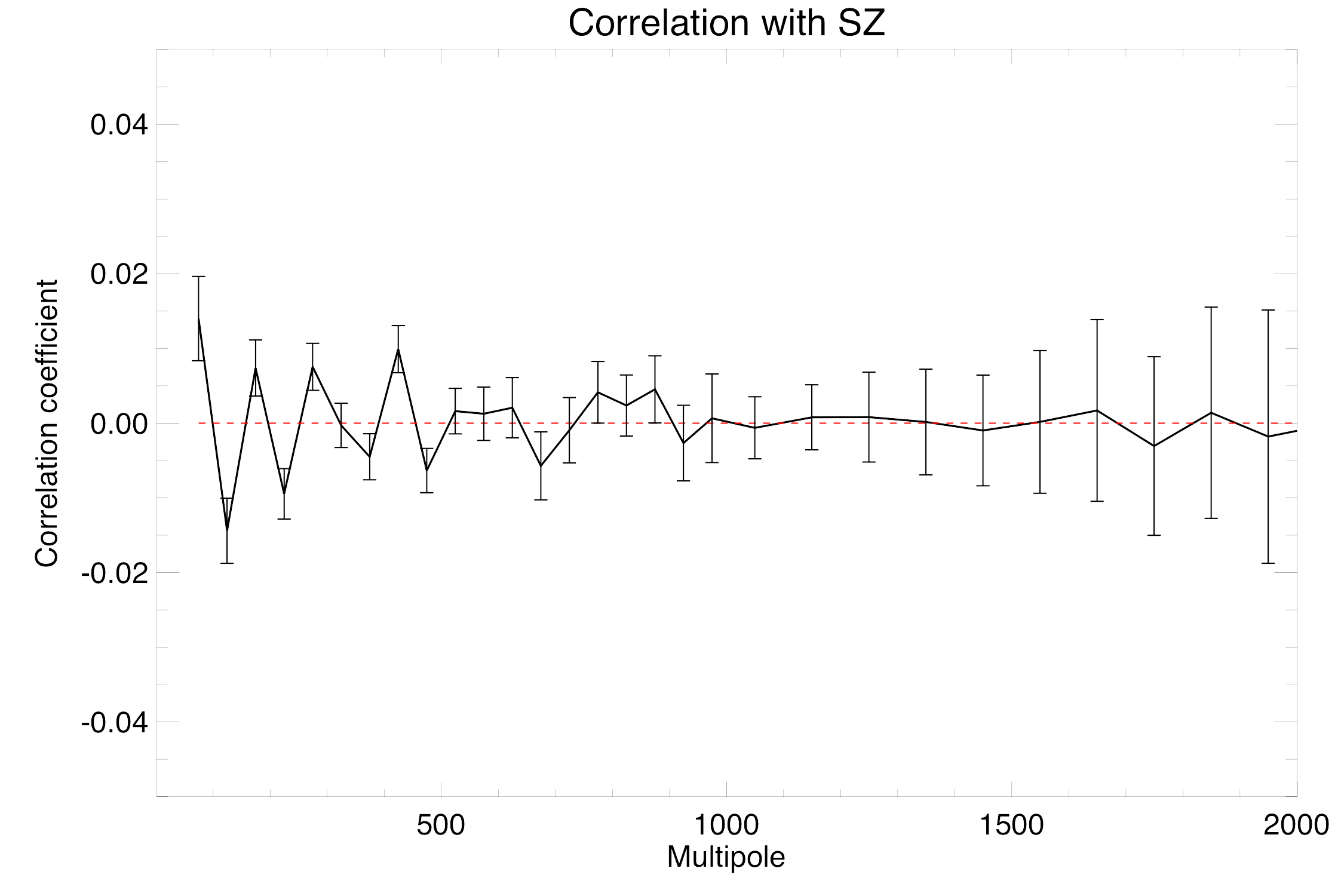} \hspace{0.2cm} 
\includegraphics[scale=0.2]{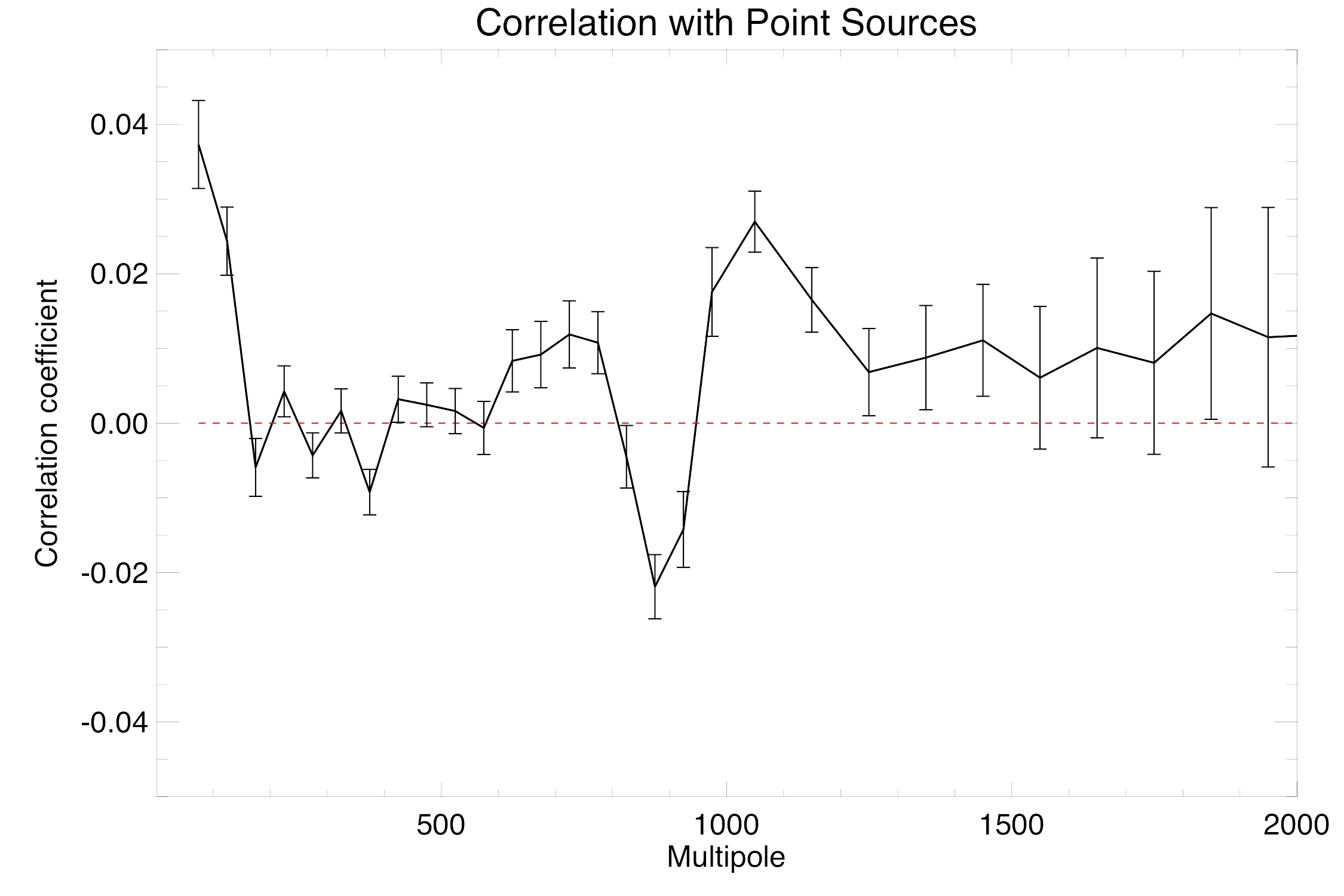}
}}
\caption{Correlation coefficient in the harmonic space of the input foregrounds with the estimated CMB map. Error bars are set to $1 \sigma$ and computed from $50$ Monte-Carlo simulations of CMB + noise.}
\label{fig_simus_xcorr}
\end{figure*}

\jbc{The recovered CMB map is displayed in Figure~\ref{fig_simus_CMB}. The error map, defined as the difference between the estimated and the input CMB maps, is shown in Figure~\ref{fig_simus_CMB_error}. For a better visualization of the foreground residuals, the error map has been downgraded to a resolution of $1.5$ degrees. The error map contains some traces of instrumental noise as well as some foreground residuals. Apart from some of the point sources at high latitudes, most foreground residuals seem to be well concentrated in the vicinity of the Galactic center, which is expected.\\
In order to evaluate the quality of our estimated CMB map, the correlation coefficient of the estimated CMB map with the input foregrounds has been calculated and is displayed in Figure~\ref{fig_simus_xcorr}. Apart from the CIB, none of the foreground components have a correlation coefficient that exceeds $0.05$. Interestingly, as conjectured from the analysis of the Planck PR1 data in Section~\ref{sec:hos}, the CIB contamination increases on small scales with the correlation coefficient reaching to about $0.3$. Note that, in this evaluation, no mask has been used in order to obtain a full-sky CMB map estimate.}

\subsection*{Assessing the uncertainty of the estimated CMB map}
\begin{figure}[htb]
\centerline{\includegraphics [scale=0.25]{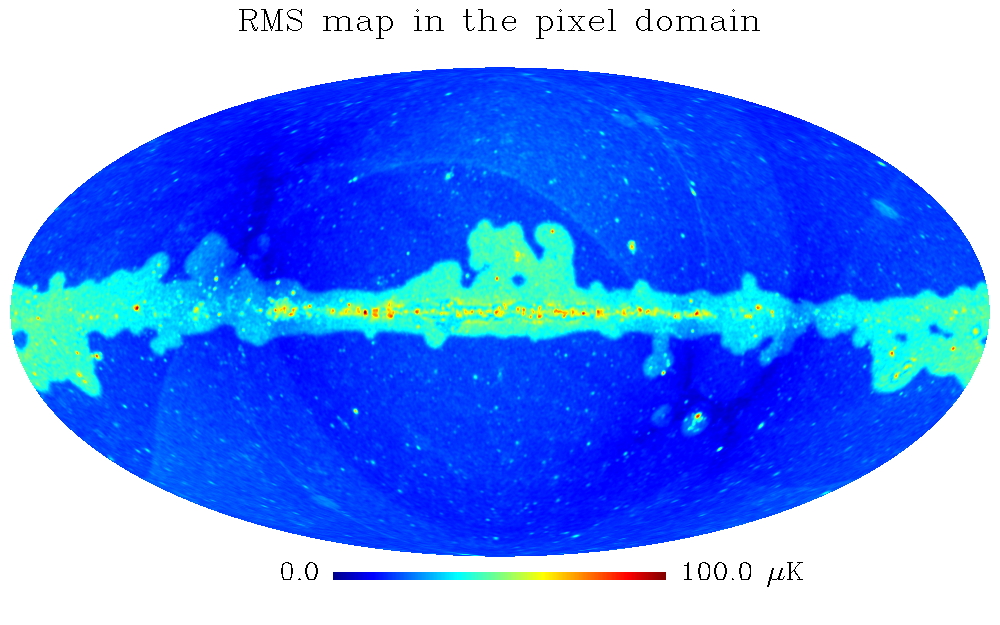}}
\caption{Uncertainty map of the estimated CMB map in the pixel domain.}
\label{fig_simus_rms_map}
\end{figure}

\begin{figure}[htb]
\centerline{\includegraphics [scale=0.2]{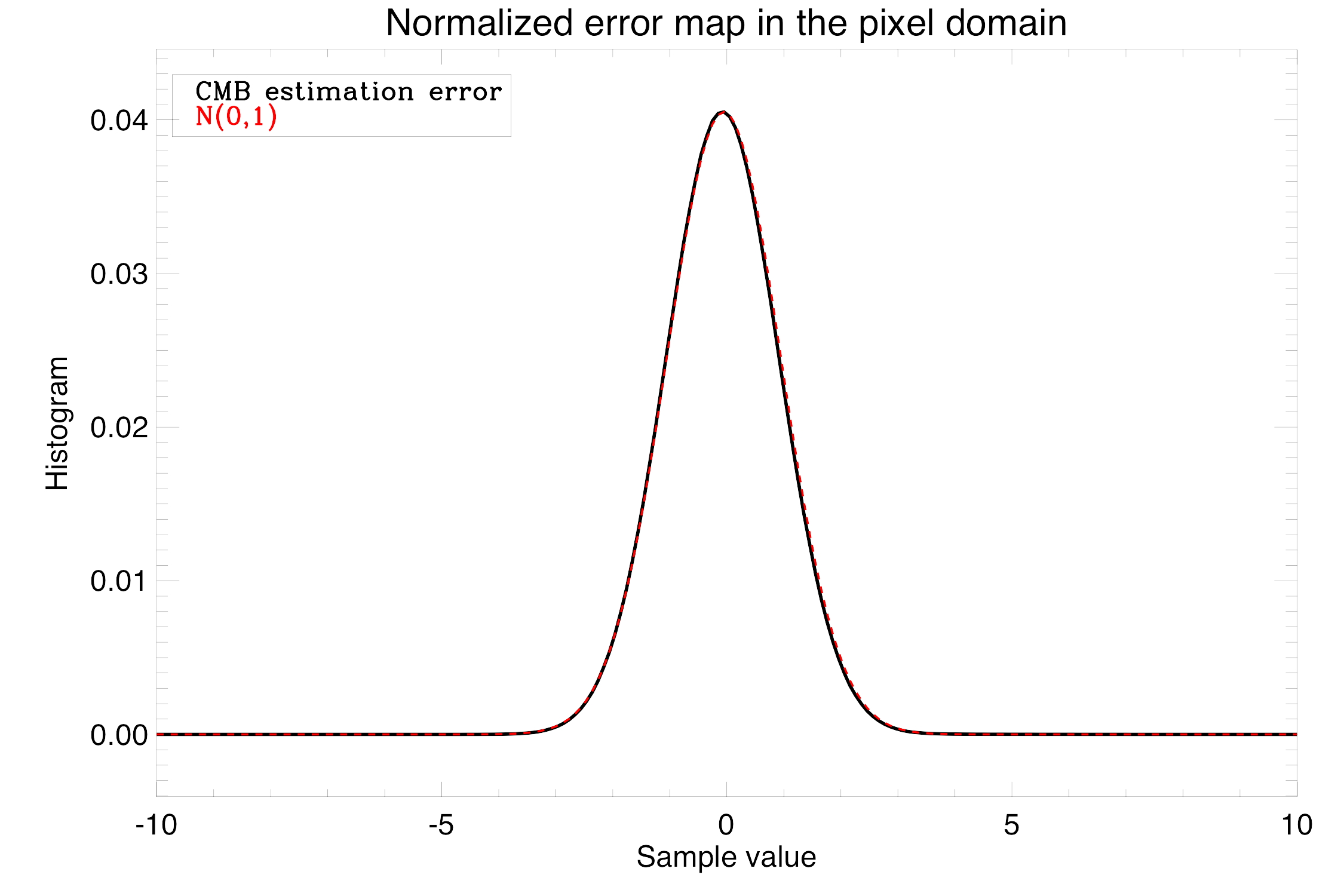}}
\caption{Histogram of the normalized error map in the pixel domain (solid black curve) and standard normal distribution (dotted red curve).}
\label{fig_simus_pix_histo}
\end{figure}

\begin{figure}[htb]
\centerline{\includegraphics [scale=0.2]{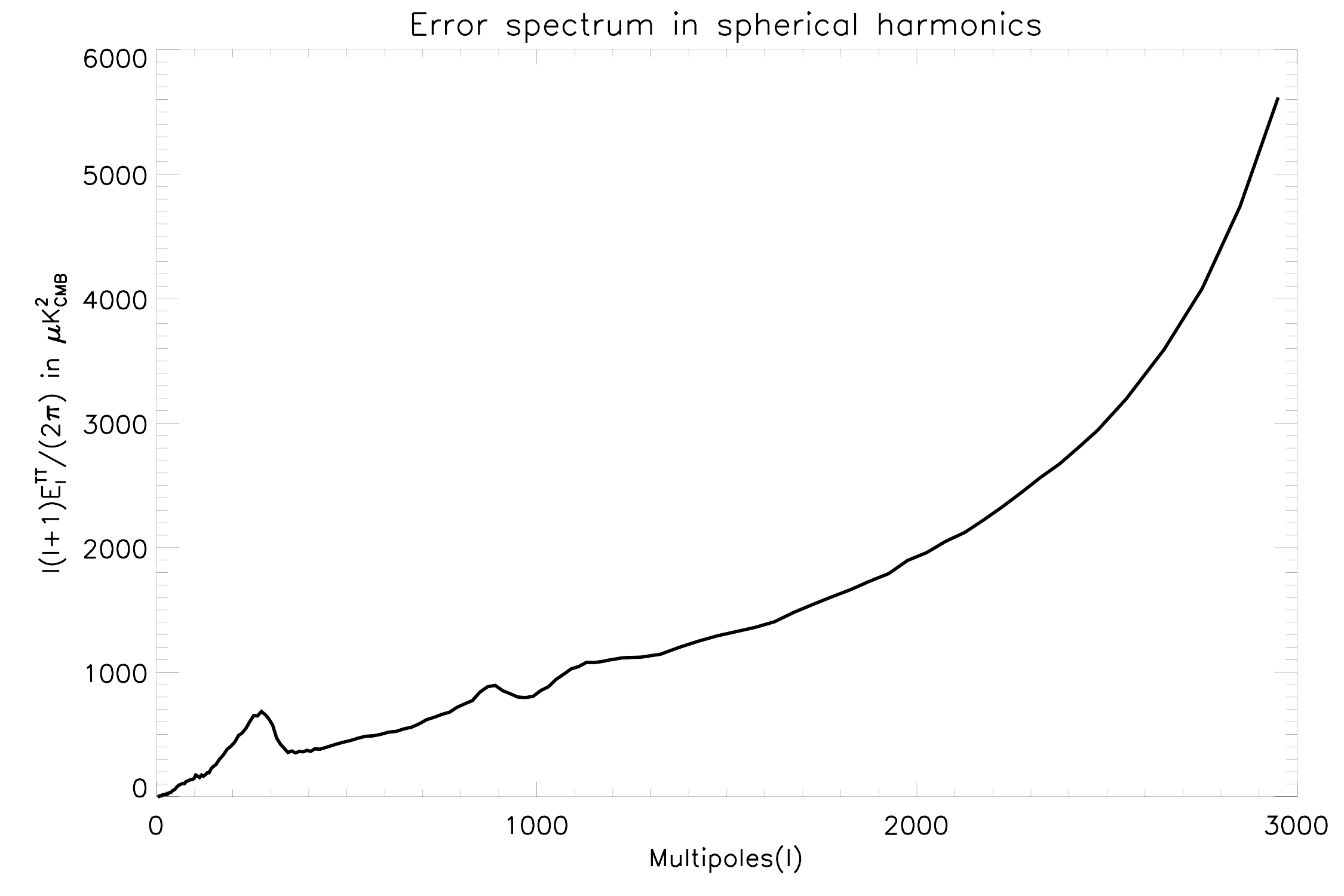}}
\caption{Uncertainty spectrum of the estimated CMB map in the harmonic domain.}
\label{fig_simus_rms_ps}
\end{figure}

\begin{figure}[htb]
\centerline{\includegraphics [scale=0.2]{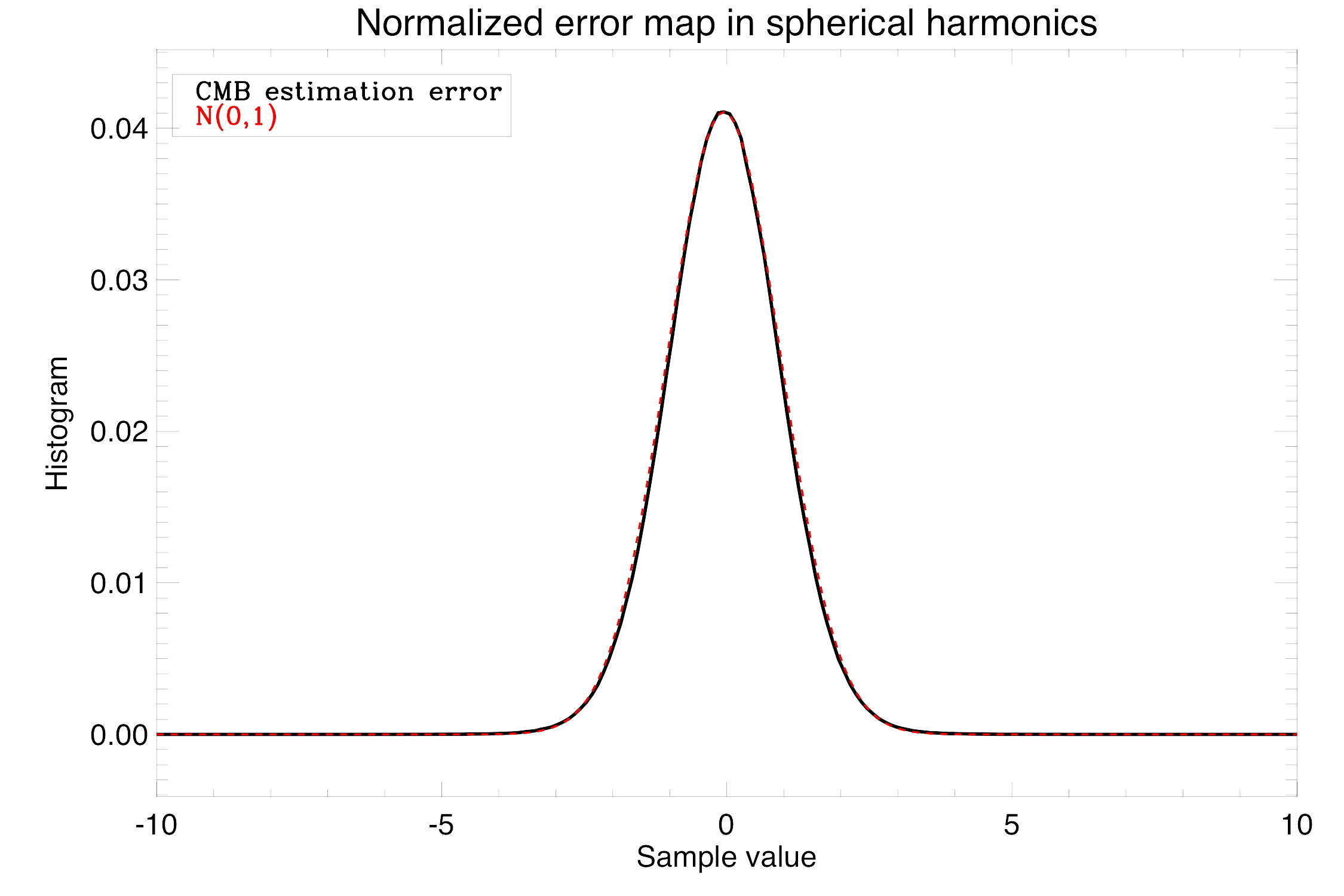}}
\caption{Histogram of the normalized error map in the harmonic domain (solid black curve) and standard normal distribution (dotted red curve).}
\label{fig_simus_sph_histo}
\end{figure}

\paragraph{\it Uncertainty estimation from simulations:}
\jbc{Even if the foregrounds have been properly removed the estimated map, we have to weigh the pixels in the map by the variance of the estimator. This would require performing Monte-Carlo simulations for all the components which compose the data, {\it i.e.} the CMB, the instrumental noise and the foregrounds, which is clearly unavailable. A more practical approach is to derive an uncertainty map which relies on measuring the variance from the error map -- displayed in Figure~\ref{fig_simus_CMB_error}.\\
Let us define $\mathcal{V}$ as the estimator variance in the pixel domain. We propose estimating $\mathcal{V}$ by the local variance of the error map. As a result, the uncertainty map, featured in Figure~\ref{fig_simus_rms_map}, has been computed by measuring the variance of the error map on overlapping patches of size $16 \times 16$ pixels. This particularly assumes that the uncertainty is stationary within each patch. As expected, with the exception of the point sources, the uncertainty map is mainly dominated by instrumental noise at high latitudes and by foreground residuals in the vicinity of the Galactic center.\\
Let us define the normalized error by:
$$
\mbox{For each pixel } k; \quad \epsilon[k] = \frac{\hat{x}[k] - x^\star[k] }{\sqrt{\mathcal{V}[k]}}\;,
$$
where $\hat{x}$ stands for the estimated CMB map and $x^\star$ is the input one. A proper estimation of $\mathcal{V}$ should be such that the normalized error $\epsilon$ asymptotically follows a Gaussian distribution with mean zero and variance one (standard normal distribution). The histogram of the normalized error is shown in Figure~\ref{fig_simus_pix_histo}; the resulting normalized error is indeed close to a standard normal distribution.\\
The foreground and the instrumental noise residuals are expected to be spatially correlated and, therefore, cannot be well characterized in the pixel domain. It can be complemented by evaluating the uncertainty in the spherical harmonics domain. Assuming that the expected uncertainty is isotropic, the CMB map uncertainty can be approximated by the power spectrum of the error map, in other words by the variance of its spherical harmonics coefficients, or $a_{\ell m}$. The resulting power spectrum $\tilde{\mathcal{V}}_{\ell}$ is shown in Figure~\ref{fig_simus_rms_ps}. Just like in the pixel domain, one can compute the normalized error in spherical harmonics:
$$
\mbox{For each } \ell, m; \quad \tilde{\epsilon}_{\ell m} = \frac{\tilde{\hat{x}}_{\ell m} - \tilde{x}^\star_{\ell m}} {\sqrt{\tilde{\mathcal{V}}_{\ell}}}\;,
$$
where $\tilde{\hat{x}}_{\ell m}$ and $\tilde{x}^\star_{\ell m}$ stand for the spherical harmonics coefficients of the estimated and the input CMB maps. The histogram of the normalized error in the spherical harmonics domain is shown in Figure~\ref{fig_simus_sph_histo}. As expected, the error distribution is close to a standard normal distribution.}

\begin{figure}[htb]
\centerline{\includegraphics [scale=0.25]{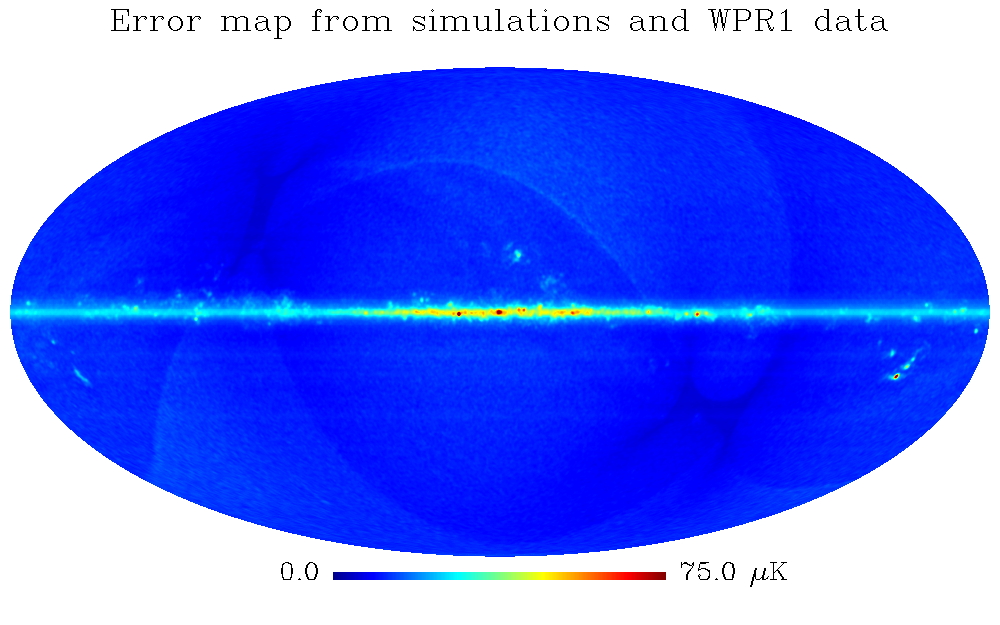}}
\caption{Uncertainty map estimated combining a level of foreground residuals estimated in bands of latitudes and the level of noise from WMAP and  Planck data.}
\label{fig_rms_lat}
\end{figure}

\begin{figure}[htb]
\centerline{\includegraphics [scale=0.25]{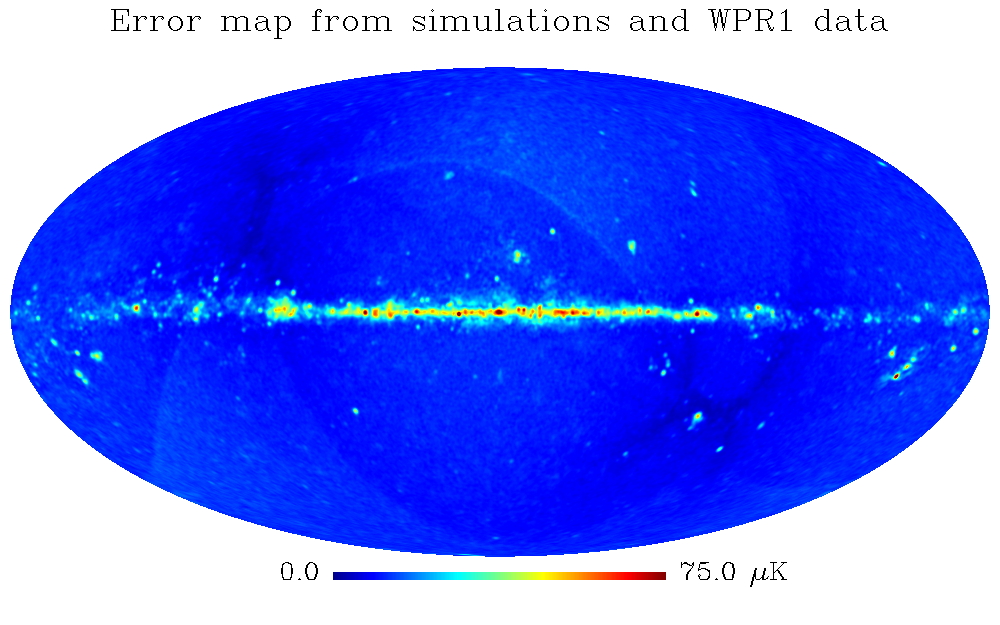}}
\caption{Uncertainty map estimated combining a level of foreground residuals at $1$ degree resolution and the level of noise from WMAP and Planck data.}
\label{fig_rms_1deg}
\end{figure}

\paragraph{\it In the case of real data:}
\label{sec:rmsmap}
\jbc{The accurate estimation of the CMB map uncertainty is very challenging in the case of real data as it is quite complex to estimate the errors at the level of CMB fluctuations or below. The quality map derived in Section \ref{sec:results} can only capture errors which are above the average noise level in the wavelet domain but is not sensitive to errors which lie below the CMB fluctuations.\\
There are two sources of error that can contaminate the CMB: i) remaining instrumental noise, ii) foreground residuals. In the case of real data, the contribution of the remaining instrumental noise to the estimate of the total error of the CMB map can be derived by computing the local variance of the half-ring difference. However, as explained previously, estimating the level of foreground residuals in the final map can only be obtained by performing simulations. In reality, simulations are only reliable at large scales where the sky has been accurately observed and studied. Therefore, we propose two approaches to estimate the uncertainty of the estimated CMB map at large scales:
\begin{itemize}
\item {\it Conservative estimate:} apart from the point source residual, the level of foreground residuals increases towards the Galactic centre. A conservative approach is to estimate the level of foreground residuals from simulations by computing the variance of the error map in bands of latitudes of $0.25$ degrees. The total error estimate, shown in Figure \ref{fig_rms_lat}, is then obtained by adding the resulting foreground residual variance and the noise variance derived from WMAP noise simulations and the Planck half ring difference. This error map provides a rough estimate of the error across latitude but is not an accurate estimate along the longitude, specifically around the Galactic centre where the error variations along the longitude is not small. This map will be made available as a product (see Section \ref{sec:repres}).\\
\item {\it  Large-scale estimate:} the latest version of the Planck Sky Model (PSM) includes physical models for the foreground components, which have been derived from the various Galactic studies. Therefore, the PSM is quite reliable at resolutions where the sky is accurately studied ({\it i.e.} resolutions greater than about $1$ degree). Hence we can estimate the level of the foreground residual by computing the local variance of the error map smoothed at $1$ degree resolution. The final error estimate, shown in Figure \ref{fig_rms_1deg}, is then obtained by adding the resulting foreground variance and the noise variance derived from WMAP noise simulations and the Planck half ring difference.
\end{itemize}
These maps provide a rough idea of the level of uncertainty of the CMB map, which also help to define reliable regions for further studies (non-Gaussianities, power spectrum estimation, etc.).}

\end{document}